\documentclass[preprintnumbers,nofootinbib,showkeys,showpacs,amsmath,amssymb]{revtex4}
\usepackage{amsmath,amssymb,graphics,epsfig,subfigure}
\usepackage{color}
\usepackage{enumitem}
\usepackage{multirow}
\usepackage{booktabs}
\usepackage{changes}
\usepackage{footnote}
\usepackage{float}

\usepackage{hyperref}
\hypersetup{colorlinks=true,linkcolor=blue,citecolor=magenta}

\begin{document}
	\renewcommand{\baselinestretch}{1.15}
	
	\title{Thermodynamics and phase transition of Bardeen-AdS-class black holes}
	
	\preprint{}

	\author{Shan-Ping Wu, Shao-Wen Wei \footnote{Corresponding author. E-mail: weishw@lzu.edu.cn}}

	\affiliation{$^{1}$Lanzhou Center for Theoretical Physics, Key Laboratory of Theoretical Physics of Gansu Province, and Key Laboratory of Quantum Theory and Applications of MoE, Lanzhou University, Lanzhou, Gansu 730000, China,\\
		$^{2}$Institute of Theoretical Physics $\&$ Research Center of Gravitation, Lanzhou University, Lanzhou 730000, China
$^{3}$ School of Physical Science and Technology, Lanzhou University, Lanzhou 730000, China}

\begin{abstract}
In a generalized parameter space, regular black holes can be regarded as non-singular solutions under specific parameters in Einstein gravity theory coupled with non-linear electromagnetic fields. Following this concept, we investigate the thermodynamic states and phase transitions of Bardeen-AdS-class black holes, revealing that the system can be classified into two categories, Type I and Type II, based on whether it adopts a pure Bardeen-AdS spacetime without event horizons or a Bardeen-AdS black hole as its phase state, each exhibiting distinct thermodynamic properties. If one includes the Bardeen-AdS black holes in the system (Type I), there will be three distinct black hole states and the phase transitions between them are analogous to the Reissner-Nordstrom-AdS black holes. On the other hand, if the pure Bardeen-AdS spacetime is included (Type II), an additional tiny black hole state emerges. A phase transition reminiscent of the Hawking-Page transition was found. Significantly, in this scenario, thermodynamical characteristic curves exhibit discontinuous behavior, which attributes to the multiple horizons. The presence of a Bardeen-AdS black hole within the phase structure of the Bardeen-AdS-class black hole profoundly modifies the thermodynamical properties of the system, highlighting the novel aspects of regular black holes.
\end{abstract}
	
	
	\maketitle
	
	
\section{Introduction}\label{Sec_Introduction}
	
	Regular black holes are a distinct solutions characterized by the absence of the center singularities. The first such model was implemented by Bardeen and is known as the Bardeen black hole~\cite{Bardeen1968}. Following it, the weak energy condition was satisfied for this regular black hole~\cite{Mars1996RBHweakcondition,Borde1996RBHtopologychange}. Unlike many conventional black holes (such as Kerr-Newman black holes), they do not violate the causality due to the absence of singularities, nor do they challenge the cosmic censorship hypothesis~\cite{Penrose1969Gravitationalcollapse}. Although the proposal of the Bardeen black hole does not directly originate from the equations of field, it has been shown that the Bardeen black hole can be accounted for within the framework of general relativity coupled with nonlinear electromagnetic fields~\cite{AyonBeato2000}. Nevertheless, this concept is general and convenient for further understanding Hayward black hole~\cite{Hayward2005Formation} and other regular black hole solutions~\cite{AyonBeato1998RegularBH,Ayon-Beato:1999NonsingularChargedBH,Ayon-Beato:1999NewRegularBH,Bronnikov:2000RegularMagneticBH,Ayon-Beato:2004FourParametricRegularBH,Hassaine:2008HighRBH,Balart:2014RBHwithElectrodynamics,FanConReBH2016,Canate2022Transforming,Li:2023RBHanalytic}. Additionally, a range of  fields can also be introduced as sources for constructing regular black holes, such as phantom fields~\cite{Bronnikov:2005RegularPhantom}, quasi-topological electromagnetic fields~\cite{Liu:2019QuasitopologicalElectromagnetism,Cisterna:2020QuasitopologicalElectromagnetism}, and the Maxwell and scalar fields in Maxwell-scalar theory~\cite{Li:2024EMSGravity}. Furthermore, various gravitational theories offer different interpretations for the regular black holes~\cite{Nicolini:2005Noncommutative,Eichhorn2022AsymptoticallySafeGravity,Ashtekar:2023LoopQuantumGravity,Nicolini2023HowStringsRBHs,Bueno:2024RBHPureGravity,Konoplya:2024DymnikovaBH,Ghosh:2022RegularizedStableKerrBH}.

After the establishment of the four laws of black hole thermodynamics, it is widely accepted that black holes are thermodynamical systems with Hawking temperature and entropy~\cite{Hawking1975ParticleCreation,Bardeen1973FourLaws,Gibbons1976ActionIntegrals}. Moreover, it seems feasible to consider black holes with nonlinear electromagnetic fields within the context of thermodynamics~\cite{Rasheed1997Zerothandfirstlaws}. The first law and Smarr relation of thermodynamics for the Einstein-Born-Infeld (-AdS, -dS) electrodynamics black hole are reasonable~\cite{Born1934Foundations,Rasheed1997Zerothandfirstlaws,Gunasekaran2012ExtendedPhaseApaceThermodynamics}. However, when one considers other regular black holes with non-linear electromagnetic fields, their thermodynamics gets some controversial issues. The requirements of the first law may lead to a conflict between the entropy-area law and the Hawking temperature~\cite{Ma2014CorrectedForm,Lan:2021Entropy,Lan:2023Entropy}, and necessitate a redefinition of the electromagnetic potential. The main reason for this situation is that the energy and magnetic charge, serving as coupling constants, appear in the action of the nonlinear electromagnetic fields. Furthermore, to address this ``coupling constant issue", numerous new forms of the first law of thermodynamics and Smarr relations have been proposed~\cite{Ma2014CorrectedForm,Azreg-Ainou:2014NoInconsistency,Zhang2016FirstLawandSmarrformula,Fan2016CriticalPhenomena,Tzikas:2018BardeenBHChemistry,Lan2021GlinerVacuum,Guo:2023RecoverThermodynamicsyRegularBlackHoles,Simovic:2023EuclideanandHamiltonianRBH,Lan:2023RBHTopicReview,Huang:2023IntermediateStates}.
	
	To clarify the reasons behind ``coupling constant issue'', let's revisit the non-linear electromagnetic part of the action
	\begin{equation}
		\mathcal{L}_{m} = \frac{12m}{ \left| g \right| ^3}\left(   \frac{\sqrt{g^2 F_{\mu \nu} F^{\mu \nu}/2}}{1+\sqrt{g^2 F_{\mu \nu} F^{\mu \nu}/2}} \right) ^{5/2},
		\label{eq_originalLm}
	\end{equation}
	used in constructing the Bardeen black hole~\cite{AyonBeato2000}. Here, the energy and magnetic charge are denoted as $m$ and $g$, respectively. Within the framework of the action, there is some unnaturalness because energy and magnetic charge are integration constants, not necessarily equivalent to the coupling constants. It implies that for the Bardeen-AdS black hole, the energy and charge cannot deviate from the coupling constants. Ultimately, in the thermodynamical discussion, we must simultaneously consider variations in energy and its associated coupling constants, which could potentially lead to inconsistencies between thermodynamical quantities and the first law of thermodynamics.
	
	To address this issue, a straightforward approach is to consider that the energy and charge are not function of the coupling constants in the Lagrangian, but rather treat $m$ and $g$ as ``real'' constants. Therefore, it becomes necessary to explore more general black hole solutions. According to the conclusions drawn in Refs.~\cite{Rasheed1997Zerothandfirstlaws,Ma2014CorrectedForm,Zhang2016FirstLawandSmarrformula,Guo:2023RecoverThermodynamicsyRegularBlackHoles}, this approach can indeed help reconcile the construction of thermodynamical quantities with the first law of thermodynamics. However, within these generalized black hole solutions, singularity avoidance occurs only under specific values of the black hole's energy and charge. This suggests that in the context of black hole thermodynamics, regular black holes may correspond to particular intermediate states~\cite{Huang:2023IntermediateStates}. Building on this premise, we intend to undertake a detailed exploration of the thermodynamics and phase transition of Bardeen black holes in the AdS background.
	
	The first crucial step is to redefine the coupling constant within the framework of nonlinear electromagnetic fields as a ``fundamental" constant, distinct from the energy or charge, thereby ensuring the internal consistency of thermodynamics. Subsequently, a class of black hole solutions with a negative cosmological constant, known as Bardeen-AdS-class black holes, is derived. In particular, we present two types of Bardeen-AdS-class black holes: one incorporating the Bardeen-AdS black hole and the other showcasing the pure Bardeen-AdS spacetime. These configurations exhibit significantly different thermodynamical properties in the issues of the black hole horizon, thermodynamical states, and phase transitions.
	
	The paper is organized as follows. In Sec.~\ref{Sec_GRcoupledNonlinearMagneticMonopole}, we provide a brief review of black hole thermodynamics within the framework of the general spherically symmetric nonlinear electromagnetic fields. In Sec.~\ref{Sec_BardeenAdSClassBHandThermodynamics} the Bardeen-AdS-class metric is obtained. The first law of thermodynamics and Smarr relation are also considered. In Sec.~\ref{Sec_PhaseStatesandPhaseTransition}, we discuss the black hole states and phase transitions for Bardeen-AdS-class black holes featuring either the Bardeen-AdS black hole or pure Bardeen-AdS spacetime. Finally, we summarize and discuss our results in Sec.~\ref{Sec_ConclusionandDiscussion}.

\section{General relativity coupled with nonlinear magnetic monopole}\label{Sec_GRcoupledNonlinearMagneticMonopole}

	As previously mentioned, the Bardeen black hole represents a specific class of regular black hole solutions within general relativity, incorporating the nonlinear magnetic monopoles. In this section, we review the general theory of gravity and examine its thermodynamical implications.

\subsection{Black hole solution}

	In Lagrangian formalism, the action for Einstein gravity minimally coupled to nonlinear electromagnetic fields is given by
	\begin{equation}
		I=\frac{1}{16 \pi }\int d^4x\sqrt{-g} \left(R+\frac{6}{L^2}-\mathcal{L}_{m}(s,a) \right),
		\label{eq_IAction}
	\end{equation}
	where $s=F_{\mu \nu} F^{\mu \nu}$, and $a$ is the parameter of the nonlinear form. The gravitational part includes the Einstein-Hilbert term and the AdS term, while the electromagnetic field part is characterized by  a nonlinear coupling $\mathcal{L}_{m}(s,a)$. By varying the action \eqref{eq_IAction} with respect to $ g_{\mu \nu} $ and $ A_{\mu} $, we derive the equations of field
	\begin{gather}
		R_{\mu \nu}-\dfrac{1}{2}g_{\mu \nu}R-\frac{3}{L^2}g_{\mu \nu}=T_{\mu \nu}, \quad T_{\mu \nu} \equiv \frac{1}{2}\left( -g_{\mu \nu}\mathcal{L} _m+4\frac{\partial \mathcal{L} _m}{\partial s}F_{\mu \lambda}{F_{\nu}}^{\lambda} \right), \label{eq_GravityEq}
		\\
		\nabla _{\mu}G^{\mu \nu}=0, \quad G^{\mu \nu} \equiv  \frac{\partial \mathcal{L}_m}{\partial s}F^{\mu \nu}.
		\label{eq_MaxwellEq}
	\end{gather}
Assuming a static and spherically symmetric spacetime, and considering the Dirac magnetic monopole, the line element can be expressed as
\begin{gather}
		ds^2=-f( r ) dt^2+\frac{1}{f( r )}dr^2+r^2d\theta ^2+r^2\sin ^2\theta d\phi ^2,
		\label{eq_MetricAnsatze}
		\\
		A = - q_m \cos\theta d\phi,
		\label{eq_MaxwellOneFormAnsatze}
	\end{gather}
Under this ansatz, Eq.~\eqref{eq_MaxwellEq} naturally holds. Furthermore, Eq. \eqref{eq_GravityEq} yields
	\begin{gather}
		f(r)=\frac{r^2}{L^2}+1-\frac{m}{r}+\frac{1}{2r}\int_r^{\infty}{dr\left( r^2\mathcal{L} _m\left( 2q_m^2/r^4,a \right) \right)}.
		\label{eq_f(r)}
	\end{gather}
	It is evident that once the electromagnetic field part of the Lagrangian $\mathcal{L}_m$ is given, $f(r)$ can be solved. Conversely, if $f(r)$ is known, $\mathcal{L}_m$ can be reconstructed~\cite{FanConReBH2016}. Note that in Eq. \eqref{eq_f(r)}, the integral with an upper limit of infinity gives us an opportunity to calculate the physical quantities directly. However, it is essential to note that the validity of Eq.~\eqref{eq_f(r)} should relies on the condition that this integral is finite.
	
	On the time-like hypersurface $\Sigma$ at $t = t_0$, the electric and magnetic charges are
	\begin{equation}
		Q_e =\frac{1}{4\pi }\int_{\partial \Sigma}{ \star G}=0, \quad Q_m = \frac{1}{4 \pi } \int_{\partial \Sigma}{F} = q_m,
		\label{eq_Qmqm}
	\end{equation}
	where $\partial\Sigma$ denotes the boundary at $r = \infty$ for hypersurface $\Sigma$, $F$ is two form given by $dA$, and $G$ is the two form $\frac{1}{2}G_{\mu \nu } dx^\mu \wedge dx^\nu$ given in Eq.~\eqref{eq_MaxwellEq}. The results indicate that the electromagnetic field $F$ is sourced by the magnetic charges. Moreover,  the difference in magnetic potential between the event horizon and infinity is represented as
	\begin{gather}
		\Phi _m= \int_{r_+}^{\infty}{dr\frac{ q_m}{r^2}\partial _s\mathcal{L} _{m}( 2q_m^2/r^4,a )}.
		\label{eq_PhimHodge}
	\end{gather}
	In asymptotically AdS spacetime, the energy can be derived from the Komar mass $-\frac{1}{8\pi }\int_{\partial \Sigma}{\star dK}$, where $K = \partial_t $ is a time-like Killing vector field. Due to the AdS background, this energy is divergent. However, by subtracting the background contribution, an effective energy can be obtained as
	\begin{equation}
		M = -\frac{1}{8\pi }\int_{\partial \Sigma}{\star dK} -M_{AdS}= \frac{m}{2}.
		\label{eq_MKomar}
	\end{equation}

\subsection{Euclidean action and thermodynamics} \label{Sec_EuclideanAction}
	
It is widely acknowledged that the thermodynamics of black holes can be derived from the partition function~\cite{Gibbons1976ActionIntegrals}
	\begin{equation}
		\mathcal{Z} = \int{\mathcal{D} g\exp \left( -I_E \right)},
	\end{equation}
For general gravitational theory, evaluating this path integral is challenging. However, using the semi-classical approach, we only consider the contributions from the on-shell solution, i.e.,
	\begin{equation}
		\mathcal{Z} \simeq \exp \left( -I_\text{on-shell} \right).
	\end{equation}
	For the gravity coupled with nonlinear electromagnetic fields, the on-shell action can be calculated by using Eqs.~\eqref{eq_MetricAnsatze}, \eqref{eq_MaxwellOneFormAnsatze}, and \eqref{eq_f(r)}. Before proceeding the calculation, it is essential to note that the singularities may occur at the event horizon in the Euclidean geometry of the black hole. Expanding the metric near the event horizon $r = r_+$, we get
	\begin{gather}
		ds^2 \simeq \rho ^2d\left( f^\prime \left( r_+ \right) t_E/2 \right) ^2+d\rho ^2+r_+^2\left( d\theta ^2+\sin ^2\theta d\phi ^2 \right),
		\\
		\rho =\frac{2}{\sqrt{f^\prime \left( r_+ \right)}} \sqrt{r-r_+}.
	\end{gather}
The cone singularity arises at $r=r_+$ unless the Euclidean time has a period $\beta =  {4\pi}/{f^\prime ( r_+ )}$. Here, $\beta$ corresponds to the black hole temperature as given by
	\begin{equation}
		T_H=\frac{1}{\beta}=\frac{f^{\prime}\left( r_+ \right)}{4\pi}=\frac{1}{4\pi r_+}\left( 1+3\frac{r_{+}^{2}}{L^2}-\frac{1}{2}r_{+}^{2}\mathcal{L} _{m}( 2q_m^2/r_{+}^{4},a ) \right) .
	\end{equation}
	Furthermore, the on-shell Euclidean action is represented by
	\begin{equation}
		I_{\text{on-shell}}=-\frac{1}{16\pi}\int_M{d^4x\sqrt{g}\left( R+\frac{6}{L^2}-\mathcal{L} _{m} \right)}-\frac{1}{8\pi }\int_{\partial M} {d^3x\sqrt{\gamma}K}+I_{ct},
		\label{eq_IEuclActiononshell}
	\end{equation}
where the first, second, and third terms are for the bulk actions of gravitational and electromagnetic fields, Gibbons-Hawking action, and the counter-term acquired through the AdS renormalization~\cite{Emparan:1999SurfaceTerms,deHaro:2000HolographicReconstruction,Skenderis:2002HolographicRenormalization}, respectively. Specifically, in four dimensions, the counter-term can be represented as
	\begin{equation}
		I_{ct} = \frac{1}{16\pi }\int_{\partial M}{d^3x\sqrt{\gamma}\left( \frac{4}{L}+LR_{\partial M} \right)}.
		\label{eq_Ict}
	\end{equation}
Combining them, we can derive the free energy as
	\begin{equation}
		F= -\frac{1}{\beta} \ln \mathcal{Z}=I_{\mathrm{on}\-\mathrm{shell}}/\beta =\frac{m}{2}-\frac{f^{\prime}(r_+)}{4\pi}\pi r_{+}^{2}.
	\end{equation}
	Subsequently, other thermodynamical quantities can be obtained by the free energy,
	\begin{gather}
		S =-\frac{\partial F}{\partial T_H}=-\frac{\partial _{r_+}F}{\partial _{r_+}T_H} = \pi r_+^2,
		\\
		M =F+T_H S = \frac{m}{2},
		\\
		\Phi _m=\frac{\partial F}{\partial Q_m}=\int_{r_+}^{\infty}{dr\frac{Q_m}{r^2}\partial _s\mathcal{L} _m(2Q_m^2/r^4,a)}.
	\end{gather}
	Clearly, the magnetic potential and energy exactly satisfy Eqs.~\eqref{eq_PhimHodge} and \eqref{eq_MKomar}. If the cosmological constants $\Lambda = -3 /L^2 $ and parameter $a$ are treated as variables, the extended first law of thermodynamics can be formulated as
	\begin{equation}
		dM=T_H dS+\Phi _mdQ_m+\mathcal{A} da+ \frac{\Theta}{8\pi }d\Lambda,
	\end{equation}
	with $\Theta = -4/3 \pi r_h^3$, and
	\begin{equation}
		\mathcal{A} =\frac{1}{16\pi }\int_{r_+}^{\infty}{d^4x\sqrt{-g}\partial _a\mathcal{L} _m(2q_{m}^{2}/r^4,a)}.
		\label{eq_A_p_aM}
	\end{equation}	
The parameter $a$ serves two main purposes: it characterizes the coupling parameter of the theory and regulates the dimension of the nonlinear electromagnetic fields. This ensures that there is a certain $\sigma$ such that the functional relation $M(\lambda^2 S, \lambda Q_m, \lambda^{-2} \Lambda, \lambda^\sigma a) = \lambda M(S, Q_m, \Lambda, a) $ holds, thereby indicating the Smarr relation,
	\begin{equation}
		M=2T_H S-2\frac{\Theta}{8\pi }\Lambda +\Phi _mQ_m+\sigma \mathcal{A} a,
	\end{equation}
	where $\sigma$ can be interpreted as the dimension of the parameter $a$. As demonstrated above, one can see that the construction of thermodynamics for black holes with general nonlinear electromagnetic fields is self-consistent. Nevertheless, this differs from the case of the Bardeen (-AdS) black hole, where the energy is related to the coupling constant in the Lagrangian, and we will discuss it in the next section.

\section{Bardeen-AdS-class black hole and thermodynamics}\label{Sec_BardeenAdSClassBHandThermodynamics}
	The Bardeen-AdS black hole can be viewed as a special solution in the model of Einstein gravity coupled with a nonlinear magnetic monopole. However, as previously mentioned, the Bardeen-AdS black hole is affected by the ``coupling constant issues''. Therefore, to avoid this problem, the Lagrangian of the nonlinear electromagnetic field shall be revised as
	\begin{equation}
		\mathcal{L}_{m} =\frac{6m_0}{q_0^3}\left(   \frac{\sqrt{q_0^2s/2}}{1+\sqrt{q_0^2s/2}} \right) ^{5/2},\ s=F_{\mu \nu} F^{\mu \nu}.
		\label{eq_NonlinearL}
	\end{equation}
	In this revision, the original $m$ and $|g|$ in Eq.~\eqref{eq_originalLm} has been replaced by $m_0/2$ and $q_0$, respectively. This action form aligns with the ``Bardeen class" for ``$\mu = 3$" in Ref.~\cite{FanConReBH2016}, while with different parameters. It should be noted that with these parameters, the physical interpretation becomes more straightforward.
	
	Using Eqs.~\eqref{eq_MaxwellOneFormAnsatze} and \eqref{eq_f(r)}, the solutions for the gravity and electromagnetic field are given by
	\begin{gather}
		f\left( r \right) =\frac{r^2}{L^2}+1-\frac{m}{r}+m_0\left( \frac{q_m}{q_0} \right) ^{3/2}\left( \frac{1}{r}-\frac{r^2}{\left( r^2+q_mq_0 \right) ^{3/2}} \right),
		\label{eq_GeBarBH}
	\end{gather}
	and
	\begin{equation}
		F = q_m \sin \theta d\theta \wedge d\phi.
	\end{equation}
The metric \eqref{eq_GeBarBH} differs from the Bardeen-AdS black hole unless $m=m_0$ and $q=q_0$. Thus, this black hole solution should be considered as an extension of the Bardeen-AdS black hole within the framework of the nonlinear electromagnetic field model. Following Ref.~\cite{FanConReBH2016}, we refer to this spacetime with event horizons as the Bardeen-AdS-class black hole. Notably, the energy and magnetic charge are given by
	\begin{equation}
		M = \frac{1}{2} m, \quad Q_m = q_m.
	\end{equation}
By expanding $f (r)$ near $r=0$, we obtain
	\begin{equation}
		f\left(r\right) = m_0\left( \left( \frac{q_m}{q_0} \right) ^{3/2}-\frac{m}{m_0} \right) \frac{1}{r}+1+r^2\left( \frac{1}{L^2}-\frac{m_0}{q_0^{3}} \right) +\mathcal{O} \left( r^3 \right).
		\label{eq_fr0}
	\end{equation}
The sign of the leading term determines whether the singularity at $r=0$ is space-like or time-like. When $(q_m / q_0) ^{3/2}=m/m_0$, the curvature singularity at $r = 0$ disappears. These changes alter the structure of spacetime, and the Penrose diagrams are presented in the Appendix~\ref{Appendix_PenroseDiagram}. Moreover, Eq.~\eqref{eq_fr0} also indicates that, at small scale, the effects of the nonlinear magnetic charge and mass counterbalance each other, potentially avoiding a spacetime singularity when in equilibrium.
	
\begin{figure}[h]
		\begin{center}
			{\includegraphics[width=4cm]{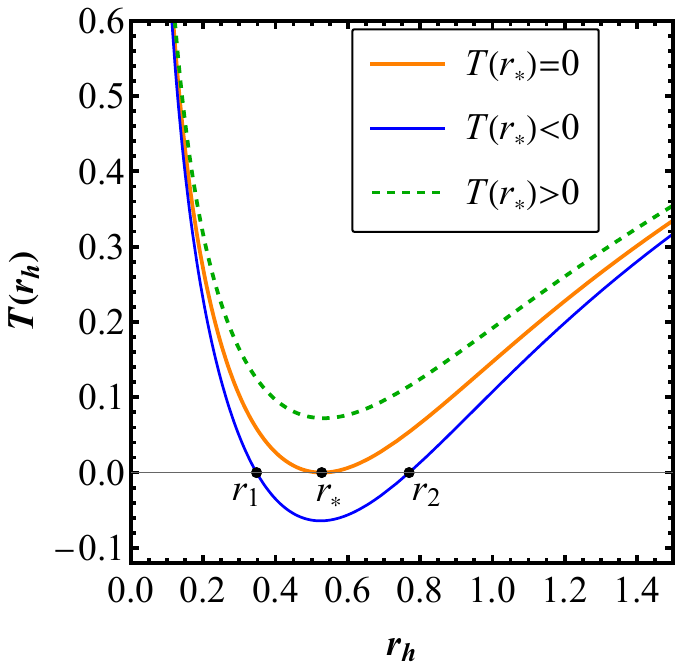}}
		\end{center}
		\caption{The sketch of the curves for $T(r_h)$. Here, we set $L=1$ and $q_m = q_0 =1$. The relationship between $r_h$ and $T$ is shown for $m_0 =4.06$, $m_0 = 5.00$, and $m_0 = 3.00$, depicted as an orange thick line, a blue thin line, and a green dashed line, respectively.}
		\label{FigrhT}
	\end{figure}

	Another noteworthy point is that the spacetime defined by metric \eqref{eq_GeBarBH} can have multiple horizons. Through a simple analysis, we confirm that the Bardeen-AdS-class black hole can have one, two, or three horizons. To illustrate it, let $r_h$ and $r_+$ denote the radii of the horizons and outer horizon (event horizon), respectively. For the black hole solution, we have
	\begin{equation}
		f\left( r_h,m \right) =0\quad \Longrightarrow \quad m=m\left( r_h \right),
		\label{eq_m(rh)}
	\end{equation}
	which relates the parameter $m$ to $r_h$. There are several important points to clarify. First, $r_h$ may not necessarily equal $r_+$ and can represent the inner horizon radius, even though both $r_h$ and $r_+$ satisfy the equation $f(r,m) = 0$. Second, for the Bardeen-AdS-class black holes, the range of $r_h$ is $(0,+\infty)$, but $r_+$ is often restricted to a smaller range. Third, for black holes characterized by a single-valued $m$ (with other parameters fixed), different values of $r_h$ may correspond to the same $m$, implying redundant representations of the black hole if one uses $r_h$ instead of $m$. However, this issue does not arise when using the outer horizon radius $r_+$. Despite this fact, discussing $r_h$ is still helpful because the range of $r_+$ can be inferred from the analysis of $r_h$. For further discussion, we introduce
	\begin{equation}
		T\left( r_h \right) =\frac{1}{4\pi}\left. \partial _rf\left( r,m\left( r_h \right) \right) \right|_{r=r_h},
		\label{eq_T(r_h)}
	\end{equation}
	which is proportional to the gradient of $f(r,m)$ at $r = r_h$. Even though the expressions of the function $T(r_h)$ and the black hole temperature $T_H$ are identical, there is a subtle difference because $r_h$ is not always the radius of the outer horizon. Some characteristic curves of $T(r_h)$ are shown in Fig.~\ref{FigrhT}, where $r_*$ is given by
	\begin{equation}
		r_*=\frac{\sqrt{q_0q_m}}{\sqrt{30}}\sqrt{-3L^2/\left( q_0q_m \right) +\sqrt{\left( 3L^2/\left( q_0q_m \right) +40 \right) 3L^2/\left( q_0q_m \right)}}.
		\label{eq_rstar}
	\end{equation}
The equation $T(r_h) = 0$ has two real roots if
	\begin{equation}
		\frac{m_0q_m}{q_{0}^{2}}\ge \frac{\left( 30-3L^2/\left( q_0q_m \right) +\sqrt{9L^4/\left( q_0q_m \right) ^2+120L^2/\left( q_0q_m \right)} \right) ^{7/2}}{225\sqrt{30}\left( -3L^2/\left( q_0q_m \right) +\sqrt{9L^4/\left( q_0q_m \right) ^2+120L^2/\left( q_0q_m \right)} \right) ^2}.
		\label{eq_ConditionT=0}
	\end{equation}
	Considering the continuity of $ f(r) $ for $r \in (0, \infty)$, we can conclude that the equation $f(r) = 0$ has at most three solutions. By substituting several specific parameter values, it is easy to find that $f(r) = 0$ can have zero, one, two, or three roots, corresponding to the number of the horizons.
	
	Although the number of horizons is known, we still need to clarify the range of $r_+$. From Fig.~\ref{FigrhT}, we ascertain the following results for the black hole solutions:
	\begin{itemize}
		\item $T(r_*) \geqslant 0$. Only one horizon exists, implying $r_+ = r_h$ and the range of $r_+$ is $(0,\infty)$.
		\item $T(r_*) < 0$. For $r_h\in [r_2, \infty)$ and $r_h \in (r_1, r_2)$, we have $r_+ = r_h$ and $r_+ \ne r_h$, respectively. For $r_h \in (0, r_1]$, if $f_2 > 0 $, $r_h  = r_+$, while $f_2 \leqslant 0 $, $r_h \ne  r_+$, where
			$f_2=\left. f\left( r,m\left( r_h \right) \right) \right|_{r=r_2}$.  As a result, the range of $r_+$ is $\left[ r_2,\infty \right) \cup \left\{ r_+ \middle| 0<r_+\le r_1\land \,\,f\left( r_2,m( r_+ ) \right) >0 \right\} $.
	\end{itemize}
It must be noted that not all $r_h$ satisfying $T(r_h) > 0$ represent the radius of the event horizon. The signs of $T(r_*)$ and $f_2$ also need further examination. By considering these aspects, we can determine whether $r_+=r_h$ holds and establish the range of $r_+$. Based on the general analysis above, the range of $r_+$ for Bardeen-AdS-class black holes may appear complex. However, this is not the case. In Sec.~\ref{Sec_PhaseStatesandPhaseTransition}, we will provide the range of $r_+$ and employ a more intuitive approach.
	
	According to the discussion in Sec.~\ref{Sec_EuclideanAction}, the Hawking temperature is determined from the geometry near the event horizon, resulting in
	\begin{equation}
		T_H=\frac{1}{4\pi}\left( \frac{1}{r_+}+3\frac{r_+}{L^2}-3m_0r_+\sqrt{\frac{q_{m}^{5}}{q_0\left( q_0q_m+r_{+}^{2} \right) ^5}} \right).
		\label{eq_HawkingTemperatureOfGBBH}
	\end{equation}
	The corresponding energy is given by
	\begin{equation}
		M = \frac{1}{2} m=\frac{1}{2}\left( \frac{r_{+}^{3}}{L^2}+r_++m_0\left( \frac{q_m}{q_0} \right) ^{3/2}\left( 1-\frac{r_{+}^{3}}{\left( r_{+}^{2}+q_0q_m \right) ^{3/2}} \right) \right).
	\end{equation}
	Furthermore, the extended first law of black hole thermodynamics is expressed as
	\begin{equation}
		dM=T_HdS+ \frac{\Theta}{8 \pi } d\Lambda+\Phi _mdQ_m+\phi _{m}dm_0+\phi _{q}dq_0,
		\label{eq_extendThermGBBH}
	\end{equation}
	where $\phi_{m}$ and $\phi_{q}$ arise from variations of  $M$ with respect to $m_0$ and $q_0$, respectively. The corresponding thermodynamical quantities for the black holes are described by
	\begin{align}
		&S=\pi r_{+}^{2},
		\\
		&\Theta= -\frac{4}{3}\pi r_{+}^{3},
		\\
		&\Phi _m =\frac{3m_0}{4q_0}\sqrt{\frac{q_m}{q_0}}\left( 1-\frac{r_{+}^{5}}{\left( r_{+}^{2}+q_0q_m \right) ^{5/2}} \right),
		\\
		&\phi _{m} =\frac{1}{2}\left( \frac{q_m}{q_0} \right) ^{3/2}\left( 1-\frac{r_{+}^{3}}{\left( r_{+}^{2}+q_0q_m \right) ^{3/2}} \right) ,
		\\
		&\phi _{q} =\frac{3m_0}{4q_0}\left( \frac{q_m}{q_0} \right) ^{3/2}\left( -1+\frac{r_{+}^{5}+2q_0q_mr_{+}^{3}}{\left( r_{+}^{2}+q_0q_m \right) ^{5/2}} \right).
	\end{align}
	It is noteworthy that both $m_0$ and $q_0$ serve as coupling constants in the Lagrangian, distinct from the integration constants $m$ and $q_m$. However, for the Bardeen-AdS black hole, $m$ and $q_m$ are equal to $m_0$ and $q_0$, respectively, leading to specific thermodynamical issues. From Eq.~\eqref{eq_extendThermGBBH}, it becomes evident that identifying $m$ with $m_0$ leads to a contradiction between the entropy-area law and the Hawking temperature, as well as inconsistencies with other thermodynamical quantities and the first law. Similarly, identifying  $q_m$ and $q_0$ may introduce inconsistencies between $\Phi_m$ and Eq.~\eqref{eq_PhimHodge}. Furthermore, setting $m = m_0$ and $q = q_0$ aligns the first law with that found in Ref.~\cite{Zhang2016FirstLawandSmarrformula,Guo:2023RecoverThermodynamicsyRegularBlackHoles}. All these observations suggest that the thermodynamics of Bardeen-AdS black holes differs from that of general nonlinear electromagnetic field models.
	
	On the other hand, directly incorporating certain integral constants ($m$ and $q_m$) as coupling constants in the action is not natural. To reproduce the Bardeen-AdS black hole, one must set $m = m_0$ and $q_m =q_0$ in the Bardeen-AdS-class metric \eqref{eq_GeBarBH}. Nonetheless, it is essential to keep in mind that $m$ relates to the energy, while $m_0$ serves as a coupling constant. Therefore, we regard Bardeen-AdS black holes as specific states where the energy $M$ of the Bardeen-AdS-class black hole precisely equals $m_0 /2$ (when $q_m = q_0$). Hence, we argue that the Bardeen-AdS black holes represent only particular states within Bardeen-AdS-class black holes, as discussed in Ref.~\cite{Huang:2023IntermediateStates}.
		
	Based on the self-consistency of the first law, the fundamental thermodynamical properties of black holes without the extension of phase space are of significant concern. To this end, we fix $L$, $m_0$, $q_0$, and focus on the ensemble with fixed $Q_m = q_0$ for simplification. Thus, the metric takes the form of
	\begin{gather}
		ds^2=-f\left( r \right) dt^2+\frac{1}{f\left( r \right)}dr^2+r^2\left( d\theta ^2+\sin ^2\theta d\phi ^2 \right), \nonumber
		\\
		f\left( r \right) =\frac{r^2}{L^2}+1-\frac{m-m_0}{r}-\frac{m_0r^2}{\left( r^2+q_{0}^{2} \right) ^{3/2}}.
	\end{gather}
	The first law of thermodynamics for the Bardeen-AdS-class black hole then reduces to
	\begin{equation}
		dM=T_HdS\quad \Longleftrightarrow \quad dF=-SdT_H,
		\label{eq_firstlaw}
	\end{equation}
	where
	\begin{equation}
		T_H=\frac{1}{4\pi}\left( \frac{1}{r_+}+3\frac{r_+}{L^2}-3m_0r_+\sqrt{\frac{q_{0}^{4}}{\left( q_{0}^{2}+r_{+}^{2} \right) ^5}} \right).
		\label{eq_THr}
	\end{equation}
It is important to note that if $Q_m$ and $q_0$ are regarded as identical, the magnetic potential becomes $\Phi_m + \phi_q$, which deviates from the conventional definition \eqref{eq_PhimHodge}. However, under this identification, the reduced first law of thermodynamics \eqref{eq_firstlaw} is self-consistent because that $Q_m$ and $q_m$ remain unchanged. In the following sections, we will consider an ensemble with fixed values of $\Lambda$, $m_0$, $q_0$ and $Q_m$ ($Q_m = q_0$), with a particular focus on phase transitions induced by temperature. This setup facilitates the discussion of black hole states and phase transitions, with intriguing results presented in Sec.~\ref{Sec_PhaseStatesandPhaseTransition}.

\section{Phase States and Phase Transitions of Bardeen-AdS-class Black Holes }\label{Sec_PhaseStatesandPhaseTransition}	
	
\begin{figure}[h]
		\begin{center}
			{\includegraphics[width=4cm]{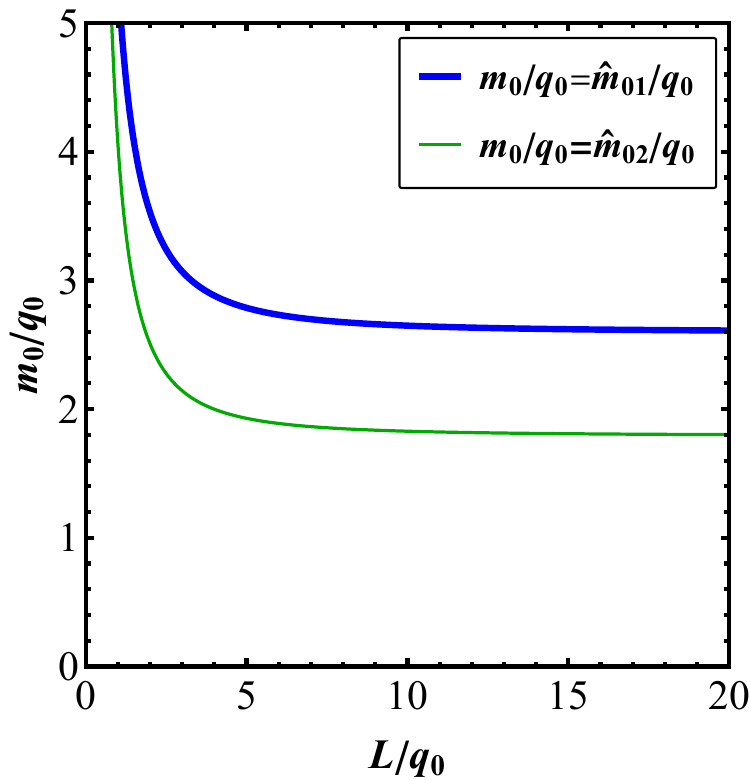}}
		\end{center}
		\caption{The parameter regions for the characteristic of the Bardeen-AdS-class black hole. The parameter space is divided into three regions by two curves. The blue bold curve and the green thin curve represent $m_0 = \hat{m}_{01}$ given by Eq.~\eqref{eq_m01expression} and $m_0 = \hat{m}_{02}$ given by Eq.~\eqref{eq_m02expression}, respectively.}
		\label{Fig_m01m02}
	\end{figure}

When $Q_m = q_0$, the Bardeen-AdS black hole emerges from the Bardeen-AdS-class black holes provided that the condition $M = m_0/2$ is satisfied. This condition highlights that Bardeen-AdS black holes can be identified by specific points (satisfying $m = m_0$) on certain thermodynamical characteristic curves of the Bardeen-AdS-class black holes. However, it is important to note that $m = m_0$ could correspond to a non-singular spacetime without event horizon, known as a pure Bardeen-AdS spacetime (state). To address this distinction, we examine the relevant equation
	\begin{equation}
		f\left(r,m_0\right) = \frac{r^2}{L^2}+1-\frac{m_0r^2}{\left( r^2+q_{0}^{2} \right) ^{3/2}} = 0.
		\label{eq_f(r,m0)eq0}
	\end{equation}
Let us first list a critical parameter
	\begin{equation}
		\frac{\hat{m}_{01}}{q_0} = \frac{\sqrt{6}\left( 6-L^2/q_{0}^{2}+\sqrt{\left( L^2/q_{0}^{2} \right) ^2+24L^2/q_{0}^{2}} \right) ^{5/2}}{9\left( -L^2/q_{0}^{2}+\sqrt{\left( L^2/q_{0}^{2} \right) ^2+24L^2/q_{0}^{2}} \right) ^2},
		\label{eq_m01expression}
	\end{equation}
	which delineates the ``boundary'' between the Bardeen-AdS black hole state and pure Bardeen-AdS state. Additionally, the relation $m_{0} = \hat{m}_{01}$ is shown in Fig.~\ref{Fig_m01m02} and Fig.~\ref{Fig_Lmbyrh}. Specifically, when $m_{0} \ge \hat{m}_{01}$, the case $m = m_0$ signifies the presence of the Bardeen-AdS black hole. Otherwise, for $m_{0} < \hat{m}_{01}$, it denotes the pure Bardeen-AdS state. In general, different values of $m_0$ give rise to two distinct types of Bardeen-AdS-class black holes. Type I contains the Bardeen-AdS black hole and type II contains the pure Bardeen-AdS spacetime. Subsequent studies will separately consider and explore the thermodynamics of these two types.
	
\subsection{Type I}\label{Sec_WithBardeenAdSBHStates}
	
\begin{figure}[h]
		\begin{center}
			{\includegraphics[width=4cm]{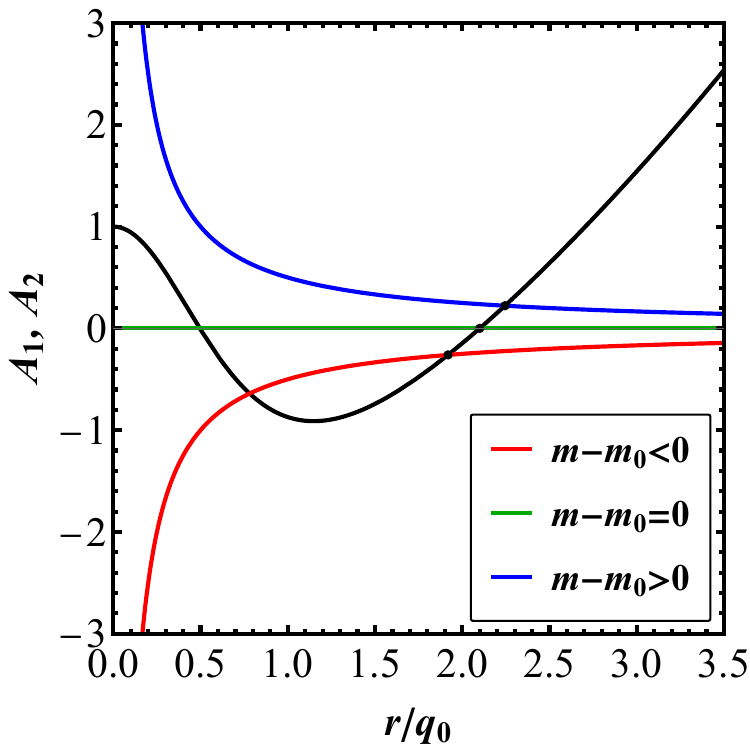}}
		\end{center}
		\caption{The sketch to characterize the number of the horizon. The black curve represents the function $A_1 (r)$. The red, green, and blue curves represent the function $A_2 (r)$ for $m-m_0 <0$, $m-m_0 = 0$, and $m-m_0 > 0$, respectively. Specifically, we set $m_0 = 6 q_0$ and $L = 2 q_0$. The red, green, and blue curves are plotted by $m = 5.5q_0$, $6q_0$, and $6.5q_0$, respectively.}
		\label{Fig_r}
	\end{figure}

To explore the thermodynamical properties of the Bardeen-AdS-class black holes, let us examine $f(r) = 0$, i.e.,
	\begin{equation}
		\frac{r_{h}^{2}}{L^2}+1-\frac{m_0r_{h}^{2}}{\left( r_{h}^{2}+q_{0}^{2} \right) ^{3/2}}= \frac{m - m_0}{r_h}.
		\label{eq_eventhorizon}
	\end{equation}
	Consequently, $r_h$ is identified by the intersection of curves $A_1(r)$ and $A_2(r)$ given by
	\begin{equation}
		A_1(r)=\frac{r^2}{L^2}+1-\frac{m_0r^2}{\left( r^2+q_{0}^{2} \right) ^{3/2}},\ A_2\left( r \right) =\frac{m-m_0}{r}.
	\end{equation}
	A schematic representation of $A_1(r)$ and $A_2(r)$ is depicted in Fig.~\ref{Fig_r}. Since the Bardeen-AdS-class black holes contain the Bardeen-AdS black holes, equation $A_1(r) = 0$ must have roots. Hence, the black curve in Fig.~\ref{Fig_r} intersect with the $r/q_0$-axis.  For $m - m_0 \neq 0$, $A_2(r)$ behaves as an inversely proportional function, suggesting that the blue curve ($m>m_0$) may have one, two, or three intersection points, while the red curve ($m<m_0$) may have zero, one, or two intersection points with the black curve. Moreover, as $m$ varies, there can be two instances where curve $A_1 (r)$ tangentially intersects curve $A_2 (r)$, indicating that $T(r_h)$ will exhibit two zero points. Through analysis, it is deduced that the maximum zero point, denoted as $r_{+m}$, also represents the minimum value of the outer horizon radius $r_+$. Consequently, the allowed range for $r_+$ can be expressed as $[r_{+m}, \infty)$. Combining with all these results, the Hawking temperature reads
		\begin{gather}
		T_H\left( r_+ \right) =\left. T\left( r_h \right) \right|_{r_h=r_+} = \frac{1}{4\pi}\left( \frac{1}{r_+}+3\frac{r_+}{L^2}-3m_0r_+\sqrt{\frac{q_{0}^{4}}{\left( q_{0}^{2}+r_{+}^{2} \right) ^5}} \right),
		r_+\ge r_{+m}.
	\end{gather}

\begin{figure}[h]
		\begin{center}
			{\includegraphics[width=4cm]{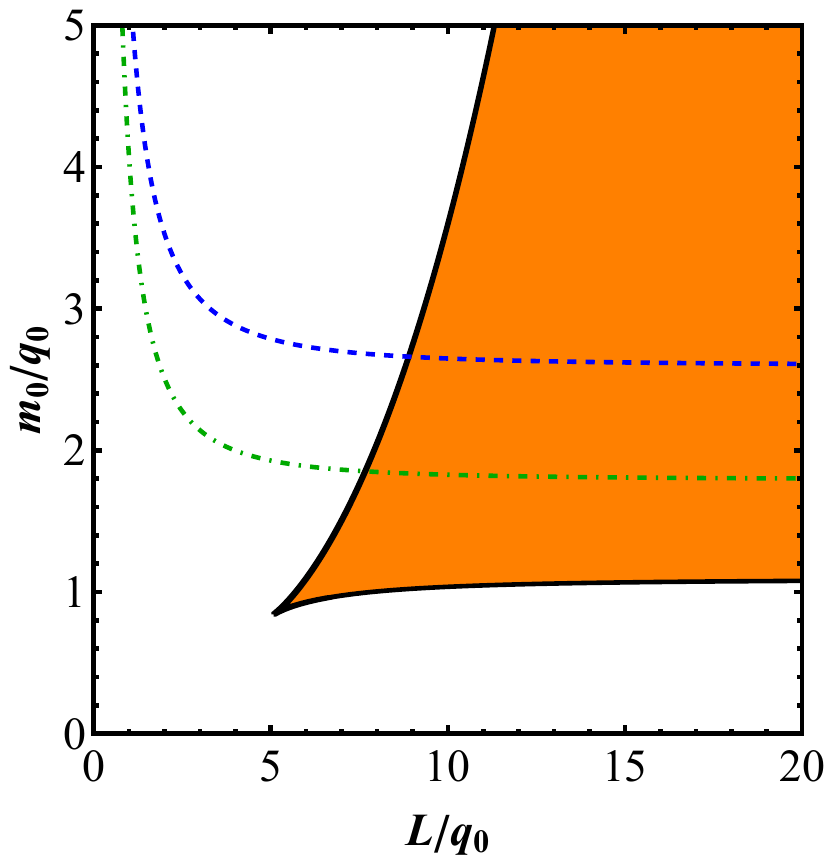}}
		\end{center}
		\caption{Parameter regions for the characteristics of $T(r_h)$. There are three stationary points for $T(r_h)$ in the orange area, and one stationary point in the white area. The dashed blue curve is plotted by $m_0 = \hat{m}_{01}$ (given by Eq.~\eqref{eq_m01expression}). The region above the dashed blue curve indicates the Bardeen-AdS black hole within the Bardeen-AdS-class black hole, while the region below represents the pure Bardeen-AdS black hole. The dot-dashed green curve is plotted for $m_0 = \hat{m}_{02}$ (given by Eq.~\eqref{eq_m02expression}). }
		\label{Fig_Lmbyrh}
	\end{figure}

For $r_+\rightarrow \infty$ and $r_+\rightarrow r_{+m}$, the asymptotic behaviors of $T_H(r_+)$ are
	\begin{align}
		& \ T_H\rightarrow \frac{3r_+}{4\pi L^2},
		\\
		&\ T_H \rightarrow \left. \partial _{r_+}T_H \right|_{r_+=r_{+m}}\left( r_+ - r_{+m} \right),
	\end{align}
respectively. Meanwhile $\partial _{r_+}T_H>0$ at $r_+=r_{+m}$ is always satisfied. This implies that the number of stable black hole phase states exceeds the number of unstable states by one. To further explore the black hole states, let us focus on the critical points determined by the following equations
	\begin{equation}
		\partial _{r_h}T\left( r_h \right) =0,\quad \partial _{r_h}^{2}T\left( r_h \right) =0,
		\label{eq_CriticalEq}
	\end{equation}
	which gives
	\begin{equation}
		 L=\frac{r_{h}^{2}\sqrt{60r_{h}^{2}-45q_{0}^{2}}}{\sqrt{2q_{0}^{4}-21q_{0}^{2}r_{h}^{2}+12r_{h}^{4}}},\quad m_0=\frac{2\left( q_{0}^{2}+r_{h}^{2} \right) ^{9/2}}{60q_{0}^{2}r_{h}^{6}-45q_{0}^{4}r_h^4}.
		\label{eq_CriticalEqReduce}
	\end{equation}
Thus, the critical relationship among parameters $L$, $m_0$, and $q_0$ is clarified, and the curve representing this relationship is plotted as the black curve in Fig.~\ref{Fig_Lmbyrh}. This curve, delineated by critical points, divides the parameter space into two regions: the orange and white regions. This partition indicates that the number of stationary points of $T(r_h)$ varies between the parameters of the orange and white regions.
	
	A potential issue arises: since the curve $r_+ - T_H$ represents only a segment of $r_h - T(r_h)$, the occurrence of critical point behavior on the $r_+ - T_H$ curve cannot be guaranteed. Nonetheless, $r_h$ satisfying Eq.~\eqref{eq_CriticalEq} must be greater than $r_{+m}$, as discussed in Appendix~\ref{Appendix_CriticalPointForTrh}. Therefore, the critical point appears exclusively on the $r_+ - T_H$ curve.
	
\begin{figure}[h]
		\begin{center}
			\subfigure[]{\includegraphics[width=4.3cm]{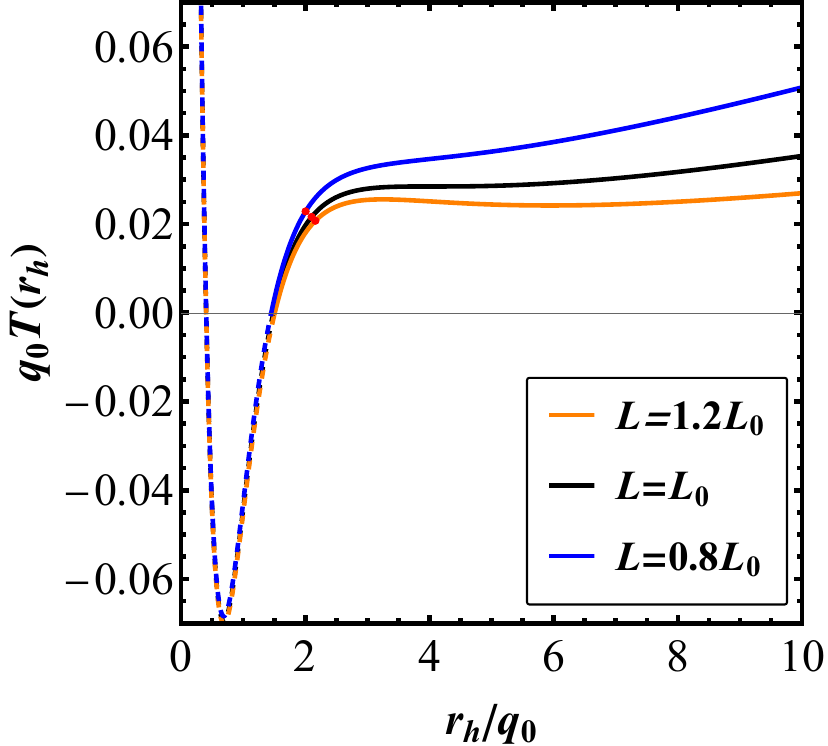}	}
			\ \ \ \ \ \
			\subfigure[]{\includegraphics[width=4.2cm]{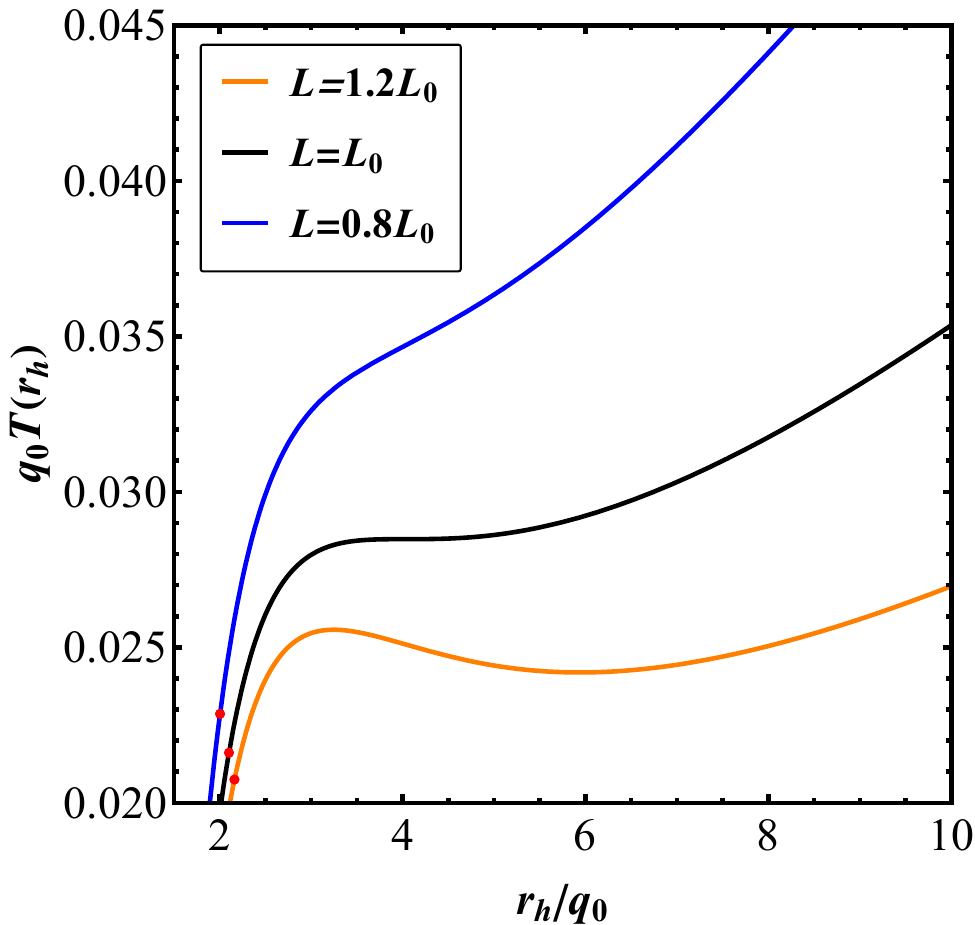} }
		\end{center}
		\caption{The $r_h - T(r_h)$ curves of Bardeen-AdS-class black hole. Here, we set $m_0 = 3q_0$ and $L_0 \equiv  9.32234 q_0$. The solid curves represent the Hawking temperature, while the dashed curves represent the ``temperature" for the inner horizon. The red point on the curve denotes the Bardeen-AdS black hole state. }\label{Fig_rhTL0}
	\end{figure}

\begin{figure}[h]
		\begin{center}
			{\includegraphics[width=4cm]{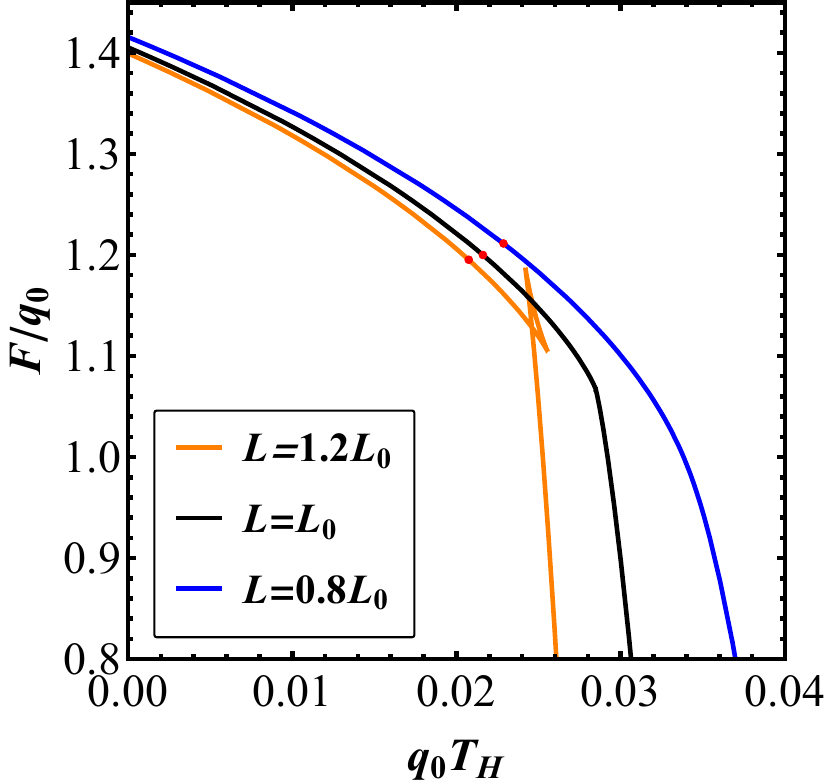}}
		\end{center}
		\caption{The $T_H - F$ curves of Bardeen-AdS-class black hole. Here, we set $m_0 = 3q_0$ and $L_0 \equiv 9.32234 q_0$. The red points on the curves denote the Bardeen-AdS black hole states.}
		\label{Fig_TFL0}
	\end{figure}

	To further examine the thermodynamical behavior within the white and orange regions in Fig.~\ref{Fig_Lmbyrh}, we set $m_0 = 3 q_0$ and vary $L$ near $L_0 \equiv 9.32234 q_0$ (parameters chosen at the boundary of these two regions in Fig.~\ref{Fig_Lmbyrh}). The corresponding curves of $r_h - T$ and $T_H - F$ are plotted in Figs.~\ref{Fig_rhTL0} and \ref{Fig_TFL0}, respectively. Only the solid curves truly represent the Hawking temperature ($r_+ - T_H$ curve), while the dashed curves represent the ``temperature" for the inner horizons. For the $r_h - T(r_h)$ curves, it demonstrates that when $L = 1.2 L_0$ (in the orange region), there are three stationary points, whereas at $L = 0.8 L_0$ (in the white region), there is only one stationary point. These observations suggest that in the orange region, the $r_h - T(r_h)$ curve exhibits three stationary points, while in the white region, there is only one stationary point. Concerning the $r_+ - T_H$ curve, the orange region corresponds to two stationary points, while the white region to none. Similar to the  Reissner-Nordstrom (RN) AdS black hole~\cite{Chamblin:1999CatastrophicHolography,Chamblin:1999ChargedAdSBH,Kubiznak:2012PVCritical,WeiLiuSci}, the Bardeen-AdS-class black hole can exist in three states: a stable small black hole, an unstable intermediate black hole, and a stable large black hole. Furthermore, as the temperature increases from zero, there is a phase transition from the small black hole state to the large black hole state.
	
\begin{figure}[h]
		\begin{center}
			\subfigure[]{\includegraphics[width=4.2cm]{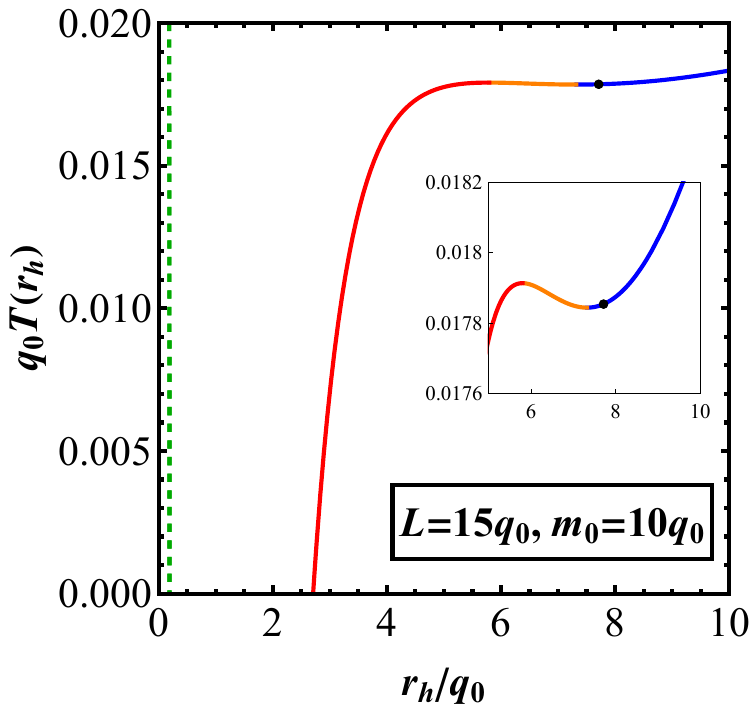}	}
			\
			\subfigure[]{\includegraphics[width=4.2cm]{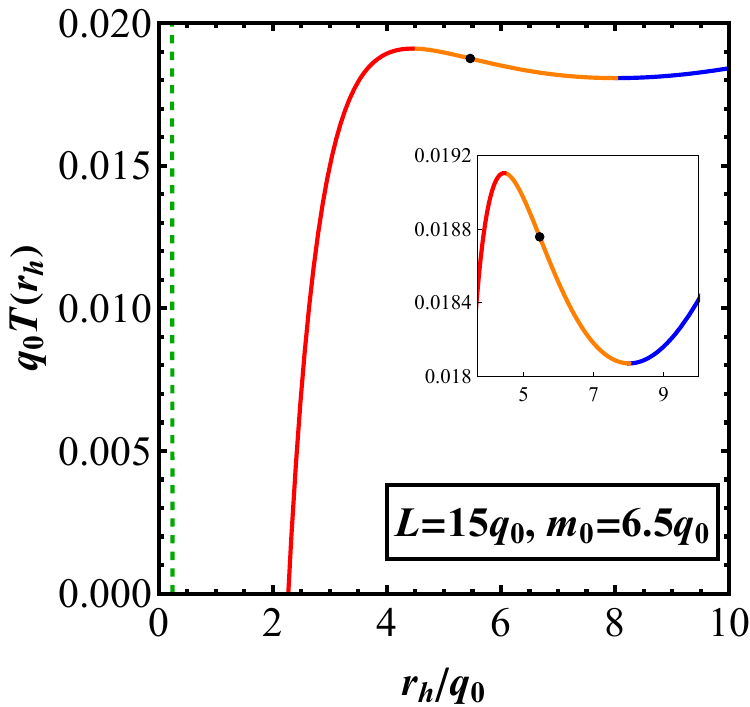} }
			\
			\subfigure[]{\includegraphics[width=4.2cm]{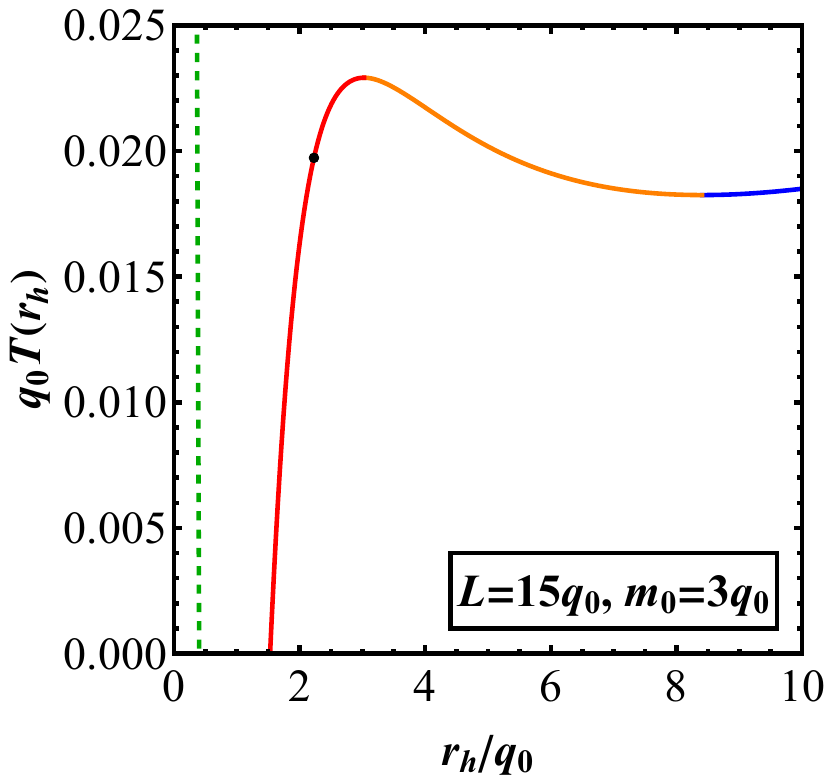} }
		\end{center}
		\caption{The $r_h - T(r_h)$ curves of Bardeen-AdS-class black hole. Here, we set $L =  15 q_0$, and plot three cases with $m_0 = 10 q_0,\  6.5 q_0 \text{ and } 3 q_0$. The solid curves denote the Hawking temperature, while the dashed curves represent the ``temperature" for the inner horizon. The black points on the curves signify the presence of the Bardeen-AdS black hole state. }\label{Figlocated}
	\end{figure}

	In Figs.~\ref{Fig_rhTL0} and \ref{Fig_TFL0}, it is evident that the Bardeen-AdS black hole state is depicted as the small black hole state. However, the Bardeen-AdS black hole can also exist in intermediate or large black hole states, as illustrated in Fig.~\ref{Figlocated}. Thus, the Bardeen-AdS black hole represents an intermediate state along the thermodynamical characteristic curves, as discussed in Ref.~\cite{Huang:2023IntermediateStates}. Moreover, in Fig.~\ref{Fig_r}, a positive correlation between $m$ and $r_+$ is observed. Consequently, it naturally follows that Bardeen-AdS black holes exist at the junction of Bardeen-AdS-class black holes with $m > m_0$ and $m<m_0$.
	
\begin{figure}[h]
 		\begin{center}
			{\includegraphics[width=4cm]{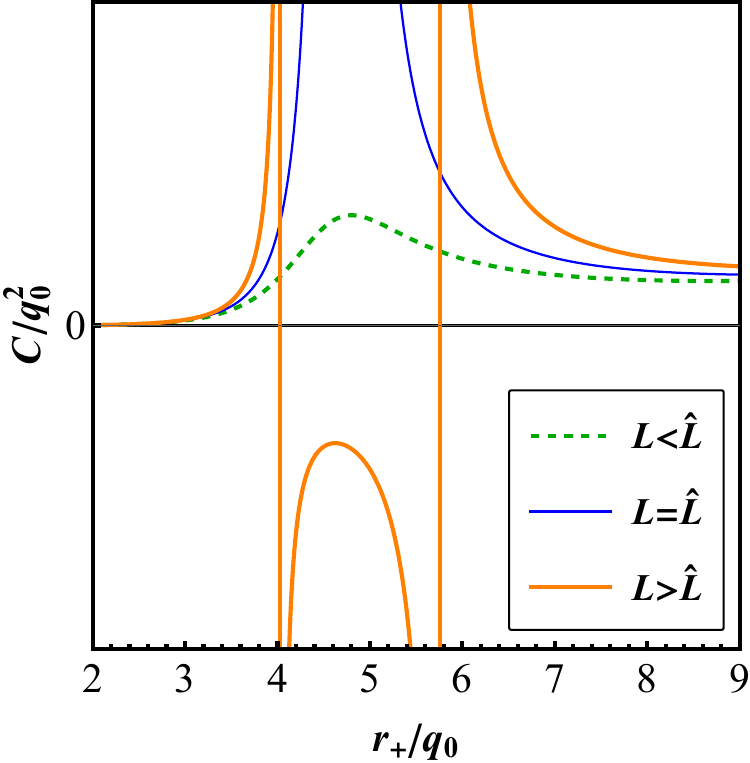}}
		\end{center}
		\caption{The heat capacity for the Bardeen-AdS black hole. Here, $\hat{L} \approx12.32 q_0$ is obtained by analyzing the denominator on the right-hand side of Eq.~\eqref{eq_rC}. Specifically, the curves corresponding to $L = 1.05 \hat{L}$, $L = \hat{L}$, and $L = 0.95 \hat{L}$ are depicted as the orange thick curve, blue thin curve, and green dashed curve, respectively.}
		\label{Fig_rpC}
	\end{figure}

	Next, let us briefly explore the thermodynamical properties of the Bardeen-AdS black hole. As an intermediate state, its thermodynamical properties should be derived from the constraints of the Bardeen-AdS-class black hole. Reduced from Eq.~\eqref{eq_THr}, the Hawking temperature of Bardeen-AdS black hole is given by
	\begin{equation}
		T_{\text{Bardeen}}=\frac{3r_{+}^{4}/L^2+r_{+}^{2}-2q_{0}^{2}}{4\pi r_+\left( q_{0}^{2}+r_{+}^{2} \right)},
	\end{equation}
	where $r_+$ is the largest solution of Eq.~\eqref{eq_f(r,m0)eq0}. In addition, $r_+$ has a lower bound $\frac{1}{6}\sqrt{-L^2+\sqrt{L^2\left( L^2+24q_{0}^{2} \right)}}$, which corresponds to the extreme Bardeen-AdS black hole. The heat capacity of the Bardeen-AdS black hole can be derived as
	\begin{equation}
		C_{\text{Bardeen}}=\left. T_H\frac{\partial S}{\partial T_H} \right|_{m_0=m\left( r_+ \right)} =\frac{2\pi r_{+}^{2}\left( r_{+}^{2}+q_{0}^{2} \right) \left( 3r_{+}^{4}+L^2r_{+}^{2}-2L^2q_{0}^{2} \right)}{3r_{+}^{6}+\left( 18q_{0}^{2}-L^2 \right) r_{+}^{4}+10L^2q_{0}^{2}r_{+}^{2}-4L^2q_{0}^{4}},
		\label{eq_rC}
	\end{equation}
	with a typical curve shown in Fig.~\ref{Fig_rpC}. Due to the inequality $T_{\text{Bardeen}}>0$, the numerator of Eq.~\eqref{eq_rC} must be non-negative, so only the denominator needs to be considered. Upon analyzing the denominator, we find that $\hat{L} = 12.32 q_0$ is a critical value. When $L \le \hat{L}$, the heat capacity is always positive, indicating stable states for the Bardeen-AdS black hole, which can only exist as small or large black hole states. Conversely, when $L > \hat{L}$, the heat capacity can be either positive or negative, depending on the radius of the Bardeen-AdS black hole. It is also important to note that the singularity of the heat capacity does not imply a phase transition of black holes. This is because in the preceding discussion of phase transitions, $m_0$ remains unchanged, while the curves in Fig.~\ref{Fig_rpC} are plotted by simultaneously varying both $m_0$ and $r_+$.
	
\subsection{Type II}\label{Sec_WithNoBardeenAdSBHStates}

	As shown above, the Bardeen-AdS black holes can be treated as intermediate states among the Bardeen-AdS-class black holes. However, for the Bardeen-AdS-class black holes with $m_0 < \hat{m}_{01}$, the pure Bardeen-AdS state will replace the Bardeen-AdS black hole and become the ``vacuum state''. In this section, we will discuss these special black hole from thermodynamical perspective. 	
	
\begin{figure}[h]
		\begin{center}
			\subfigure[]{\includegraphics[width=4.2cm]{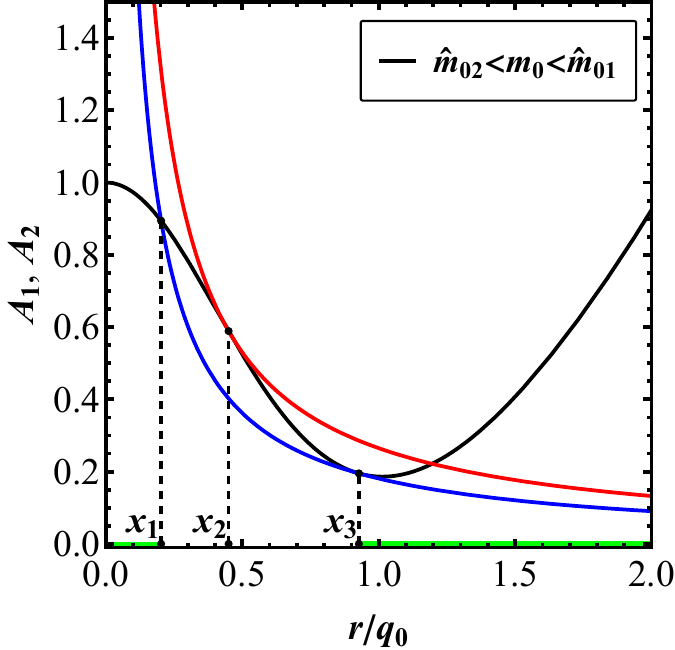}}
			\
			\subfigure[]{\includegraphics[width=4.2cm]{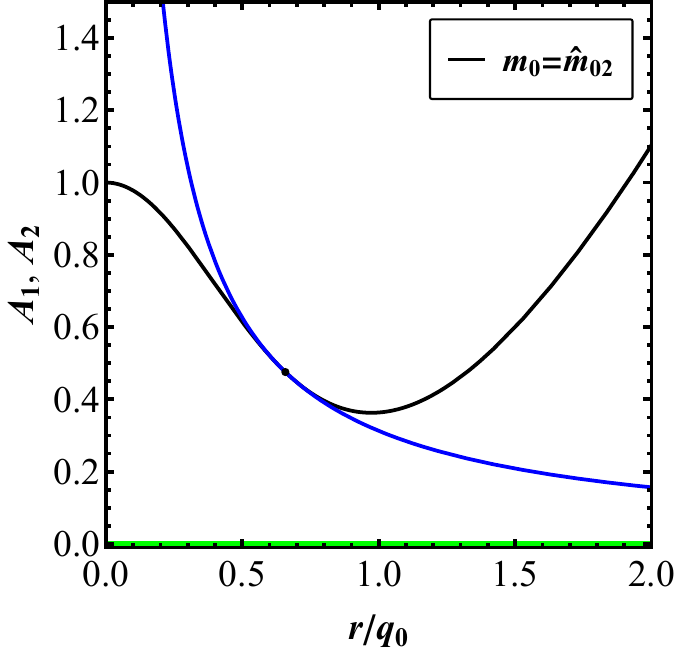} }
			\
			\subfigure[]{\includegraphics[width=4.2cm]{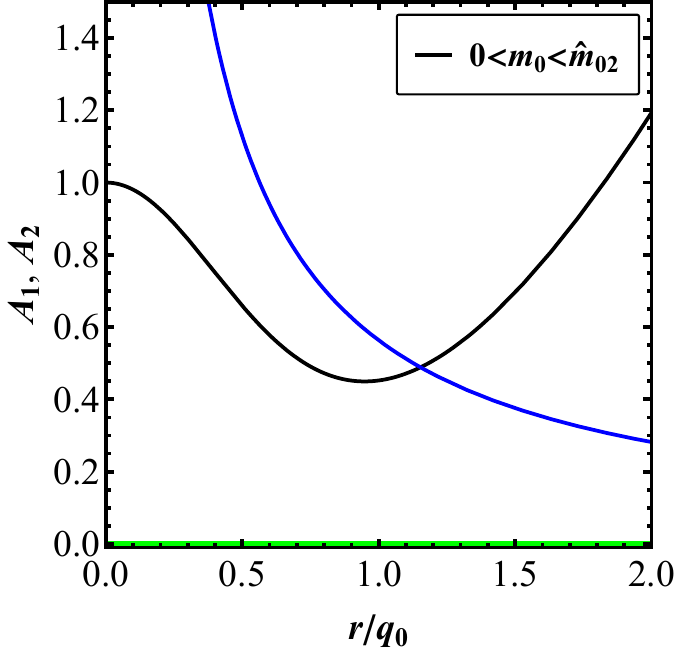} }
		\end{center}
		\caption{The sketch to characterize the number of horizon. The black curves represent the function $A_1(r)$, while the red and blue curves represent the function $A_2(r)$. Specifically, we set $L_0 = 2 q_0$, and we can derive $\hat{m}_{02} \approx 2.51 q_0 $ and $\hat{m}_{01} \approx 3.53 q_0 $ by Eqs.~\eqref{eq_m01expression} and \eqref{eq_m02expression}.
		(a) $m_0 = 1.2 \hat{m}_{02}$. We plot the black curve for $A_1(r)$, and the red and blue curves for $A_2(r)$ corresponding to $ m \approx 3.27 q_0 $ and $ m \approx 3.19 q_0$, respectively.
		(b) $m_0 = \hat{m}_{02}$. The black curve is for $A_1(r)$, and the blue curve is for $A_2(r)$ with $ m \approx 2.82 q_0$.
		(c) $m_0 = 0.9 \hat{m}_{02}$. The black curve is for $A_1(r)$, and the blue curve is for $A_2(r)$ with $ m \approx 2.82 q_0 $. Additionally, the green lines along the axis represents the allowable range for $r_+$. }\label{Fig_rN}
	\end{figure}

	First, to determine the range of $r_+$, we need to examine the horizon of the black hole. This requires studying the solution of Eq.~\eqref{eq_eventhorizon} or $A_1(r)=A_2(r)$. The Bardeen-AdS black hole state does not belong to any state within the Bardeen-AdS-class black hole, implying that the equation $A_1(r) = 0$ has no roots. This ultimately defines the parameter range of $m_0$ as $m_0< \hat{m}_{01}$. In this case, by solving the equations $A_1(r) = A_2(r)$, $\partial_r A_1(r) =\partial_r A_2(r)$, and $\partial_r^2 A_1(r) =\partial_r^2 A_2(r)$, we identify another critical value $\hat{m}_{02}$, which is expressed as
	\begin{equation}
		\frac{\hat{m}_{02}}{q_0}=\frac{\left( 30-3L^2/q_{0}^{2}+\sqrt{3L^2/q_{0}^{2}\left( 3L^2/q_{0}^{2}+40 \right)} \right) ^{7/2}}{225\sqrt{30}\left( -3L^2/q_{0}^{2}+\sqrt{3L^2/q_{0}^{2}\left( 3L^2/q_{0}^{2}+40 \right)} \right) ^2}.
		\label{eq_m02expression}
	\end{equation}
	The relations $m_{0} = \hat{m}_{01}$ and $m_{0} = \hat{m}_{02}$ are shown in Fig.~\ref{Fig_m01m02} and Fig.~\ref{Fig_Lmbyrh}. Utilizing these two critical values, $\hat{m}_{01}$ and $\hat{m}_{02}$, we provide schematic diagrams of the event horizon shown in Fig.~\ref{Fig_rN}. The absence of intersections between the black curve for $A_1(r)$ and the $r/q_0$-axis suggests that the behavior of the horizon is quite complex. However, some conclusions are clear: only when $m>m_0$, the spacetime solution is a black hole, and the horizon requires further discussion; when $m=m_0$, the spacetime manifests as a non-singular state, i.e., pure Bardeen-AdS state; when $m<m_0$, naked singularities exist in the spacetime, which we will not consider.
	
	Let us turn to the situation of the horizon for the Bardeen-AdS-class black hole shown in Fig.~\ref{Fig_rN}. In the case of $\hat{m}_{02} < m_0 < \hat{m}_{01}$, this black hole can exhibit one, two, or three horizons. Importantly, the radius range for the outer event horizon is $(0, x_1 q_0) \cup (x_3 q_0, \infty )$ rather than $(0, \infty)$, where $x_1$ and $x_3$ are the intersection and tangent points of the blue curve and the black curve, respectively. Conversely, in the case of $ 0 < m_0 \le \hat{m}_{02} $, this black hole can only possess a single horizon, with $r_+$ ranging from 0 to $\infty$. Given these distinctions, it is possible to address these two scenarios for black hole thermodynamics separately.

\begin{itemize}[leftmargin=*]
	\item \textbf{Case 1}: $\hat{m}_{02} < m_0 < \hat{m}_{01}$.
	 Not all values of $r_h \in (0, \infty)$  correspond to the radius of the outer horizon. The range of $r_+$ is determined by the intersection points of the blue and black curves, specifically $(0, x_1 q_0)\cup(x_3 q_0, +\infty)$. For $r_+ \in (0, x_1 q_0)$, the black hole temperature monotonically decreases with $r_+$, as discussed in Appendix~\ref{Appendix_CriticalPointForTrh}; for $r_+ \in (x_3 q_0, +\infty)$, the temperature exhibits the asymptotic behaviors described by
	\begin{align}
		& \ T_H\rightarrow \frac{3r_+}{4\pi L^2} \ \text{and} \  \partial _{r_+}T_H>0,
		\\
		&\ T_H \rightarrow \left. \partial _{r_+}T_H \right|_{r_+=x_3 q_0}\left( r_+ -x_3 q_0 \right) \ \text{and} \  \partial _{r_+}T_H>0,
	\end{align}
as $ \ r_+\rightarrow \infty$ or $x_3 q_0$, respectively.
Thus, for $r_+ \ge x_3 q_0$, the number of stable black hole states exceeds that of unstable states by one, contingent upon the critical points detailed in Eq.~\eqref{eq_CriticalEq}.	
The $r_h - T(r_h)$ curves are illustrated in Fig.~\ref{FigrTTFm2m1}, revealing four distinct states: the unstable tiny black hole state (red curve), the stable small black hole state (orange curve), the unstable intermediate black hole state (green curve), and the stable large black hole state (blue curve). Additionally, there exists a pure Bardeen-AdS state without an event horizon, characterized by spacetime structure given in
	\begin{equation}
		ds^2=-\left( \frac{r^2}{L^2}+1-\frac{m_0r^2}{\left( r^2+q_{0}^{2} \right) ^{3/2}} \right) dt^2+\left( \frac{r^2}{L^2}+1-\frac{m_0r^2}{\left( r^2+q_{0}^{2} \right) ^{3/2}} \right) ^{-1}dr^2+r^2\left( d\theta ^2+\sin ^2\theta d\phi ^2 \right).
		\label{eq_BardeenAdSvacuum}
	\end{equation}
	By performing a Euclidean on-shell integration, the free energy of the pure Bardeen-AdS state is determined as
	\begin{equation}
		F_{\text{Bardeen} }= \frac{1}{2} m_0.
	\end{equation}
	Notably, as $m_0$ approaches zero, the system reduces to the pure AdS spacetime with the free energy tending to zero. The difference in the free energy between the Bardeen-AdS-class black hole state and the pure Bardeen-AdS state is expressed
	\begin{equation}
		\Delta F=\frac{1}{4}\left( r_+-\frac{r_{+}^{3}}{L^2}-\frac{m_0r_{+}^{3}\left( 2r_{+}^{2}-q_{0}^{2} \right)}{\left( r_{+}^{2}+q_{0}^{2} \right) ^{5/2}} \right).
	\end{equation}
Here,  $\Delta F = 0$ signifies a possible phase transition between the pure Bardeen-AdS state and the black hole state. Additionally, several phase transition diagrams are presented in Fig.~\ref{FigrTTFm2m1}. As the temperature increases from zero, two types of phase transitions are observed: one similar to the Hawking-Page transition from the pure Bardeen-AdS state to the black hole state, and another resembling the transition from small to large black hole states, similar to RN-AdS black holes. Notably, at zero temperature, there are no the unstable tiny black hole states. Additionally, the points for zero temperature on the dashed segment indicate the coincidence of two inner horizons.
	\begin{figure}[h]
		\begin{center}
			\subfigure[ $L = 10 q_0$, $m_0 \approx 2.05 q_0$]{\includegraphics[width=4.2cm]{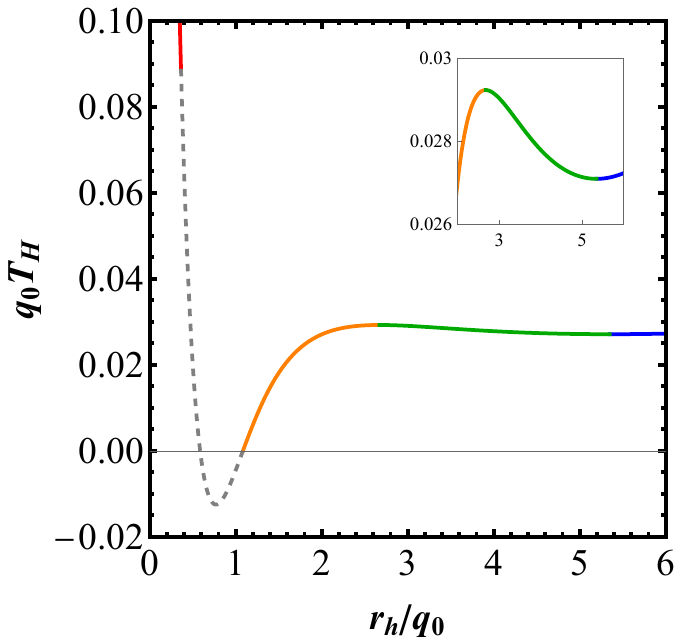}	}
			\
			\subfigure[ $L = 8 q_0$, $m_0 \approx 2.05 q_0$]{\includegraphics[width=4.2cm]{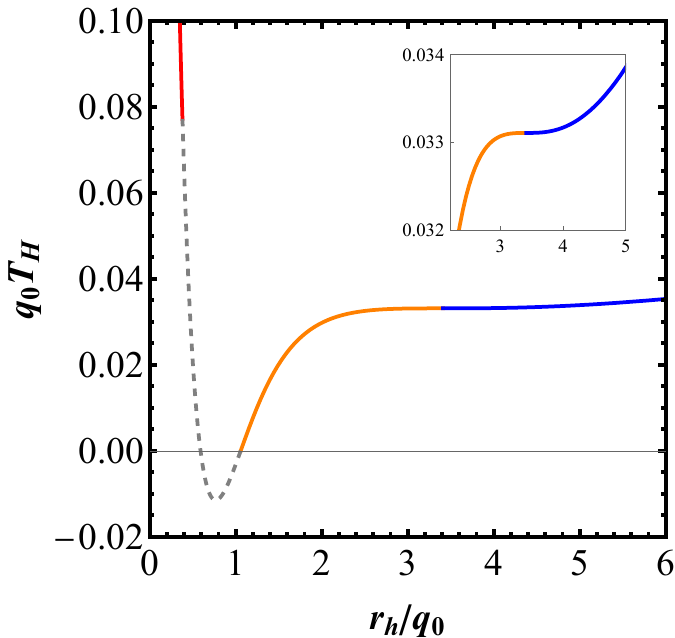} }
			\
			\subfigure[ $L = 7 q_0$, $m_0 \approx 2.05 q_0$]{\includegraphics[width=4.2cm]{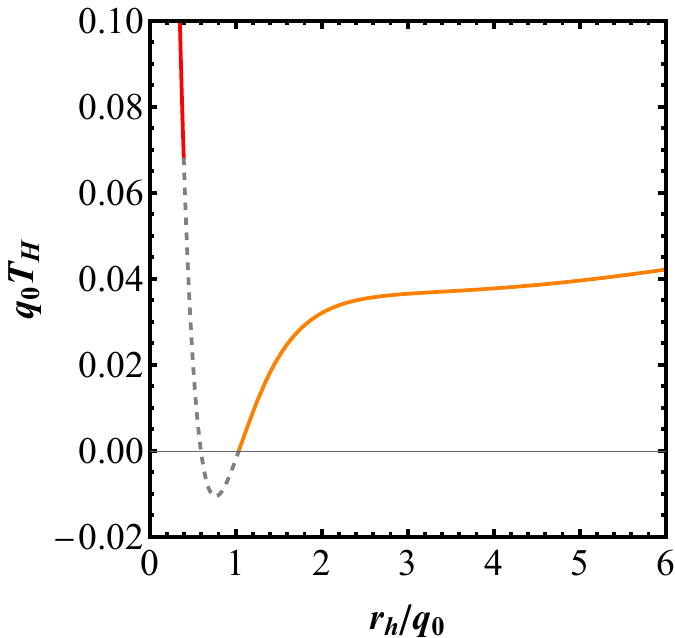} }
			\\
			\subfigure[ $L = 10 q_0$, $m_0 \approx 2.05 q_0$]{\includegraphics[width=4.2cm]{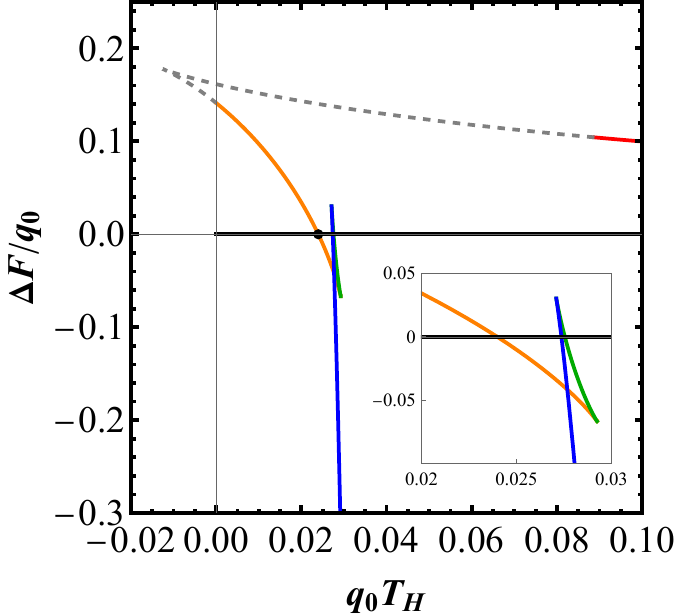}	}
			\
			\subfigure[ $L = 8 q_0$, $m_0 \approx 2.05 q_0$]{\includegraphics[width=4.2cm]{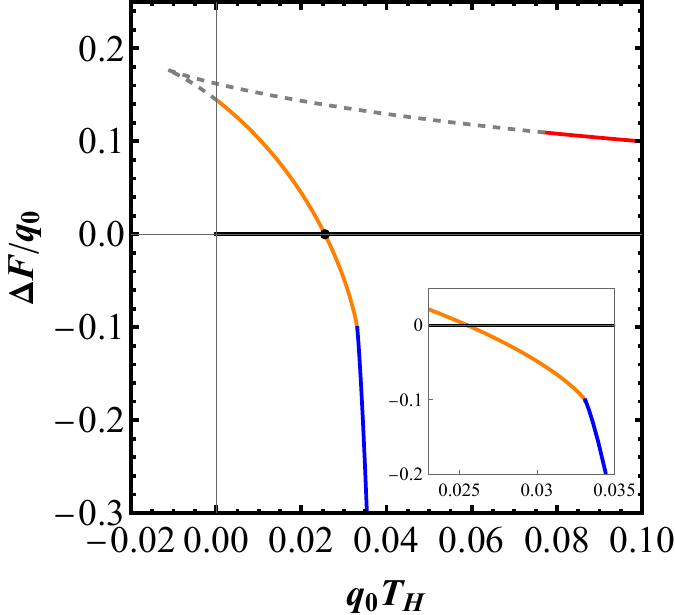} }
			\
			\subfigure[ $L = 7 q_0$, $m_0 \approx 2.05 q_0$]{\includegraphics[width=4.2cm]{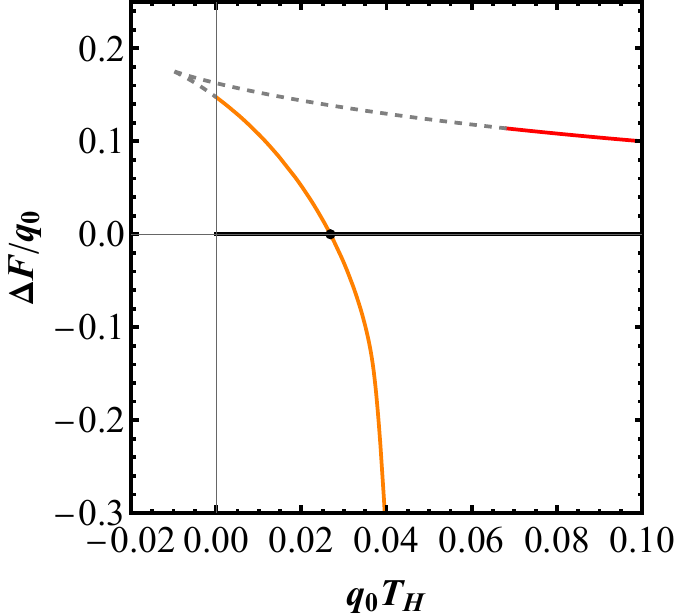} }
		\end{center}
		\caption{The characteristic curve ($r_h - T_H$ and $T_H - \Delta F$) of the Bardeen-AdS-class black hole phase states. The solid curves represent the relationship between $r_+$ and the Hawking temperature, while the dashed curves are just the $T (r_h)$ functions. The red, orange, green, and blue curves represent the tiny, small, medium, and large black hole states, respectively. The black curves represent the Bardeen-AdS states. Here, the parameters are $L = 8 q_0$ and $m_0 \approx 2.05 q_0$ satisfying Eq.~\eqref{eq_CriticalEq} for critical points.}
		\label{FigrTTFm2m1}
	\end{figure}
	
	\item \textbf{Case 2}: $0 < m_0 \le \hat{m}_{02}$.
	Bardeen-AdS-class black holes possess only one horizon, thus the range of $r_+ = r_h$ is $(0, \infty)$. The boundary behaviors of $T_H$ are described by
	\begin{align}
		&\ T_H\rightarrow \frac{3r_+}{4\pi L^2} \ \text{and} \  \partial _{r_+}T_H>0,
	\\
		&\ T_H\rightarrow \frac{1}{4\pi r_+} \ \text{and} \ \partial _{r_+}T_H=-\frac{1}{4\pi r_{+}^{2}},
	\end{align}
	for $\ r_+\rightarrow \infty$ and $0^+$, respectively. This indicates that the number of stable and unstable black hole states is equal. The specific number of states depends on Eq.~\eqref{eq_CriticalEq} for the critical points. Similar to the case $ \hat{m}_{02} < m_0 < \hat{m}_{01}$, there can be up to four black hole states and one pure Bardeen-AdS state. Utilizing Eq.~\eqref{eq_CriticalEq}, we find that the number of stationary points of the function $ T_H(r_+) $ can be one, two, or three. The temperature and free energy are displayed in Fig.~\ref{FigrTTF0m2}.
	
	Unlike the situation where $ \hat{m}_{02} < m_0 < \hat{m}_{01} $, the $ r_+ - T_H $ and $ T_H - \Delta F $ curves are continuous and complete. This is because, in this scenario, $ r_h $ and $ r_+ $ are entirely consistent. Furthermore, at the threshold point $ m_0 = \hat{m}_{02} $, both the tiny black hole phase and the small black hole phase emerge at zero temperature. Moreover, in this scenario, we have $T(r_h) = \partial_{r_h} T(r_h)=0$, implying the overlap of the black hole's three horizons. Additionally, when $m_0$ is strictly less than $\hat{m}_{02}$, the black hole state will only be excited from finite temperatures.
	
	\begin{figure}[h]
		\begin{center}
			\subfigure[$L \approx 7.67 q_0$, $m_0 \approx 1.39 q_0$]{\includegraphics[width=4.2cm]{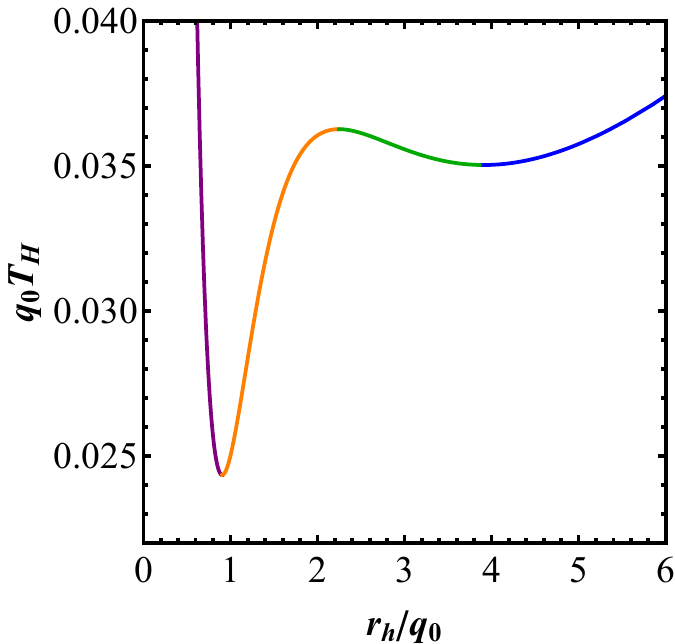}	}
			\
			\subfigure[$L \approx 7.67 q_0$, $m_0 \approx 1.85 q_0$]{\includegraphics[width=4.2cm]{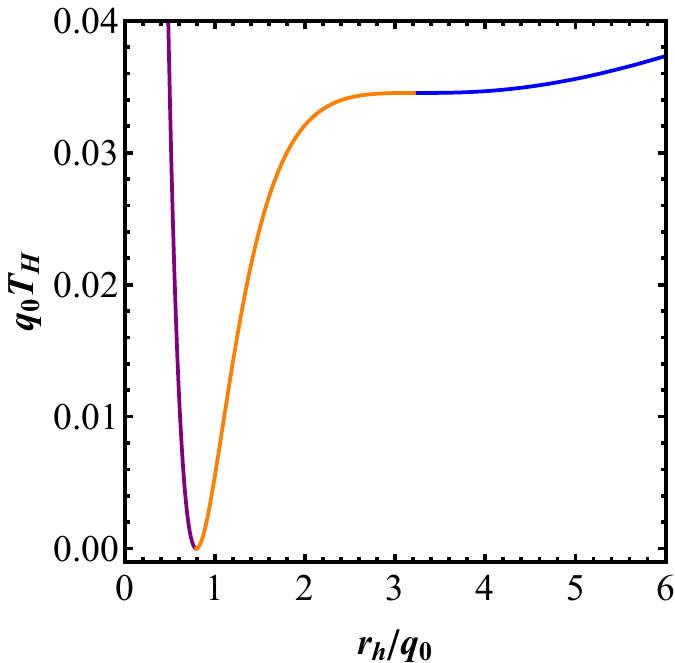} }
			\
			\subfigure[$L \approx 7.67 q_0$, $m_0 \approx 1.00 q_0$]{\includegraphics[width=4.2cm]{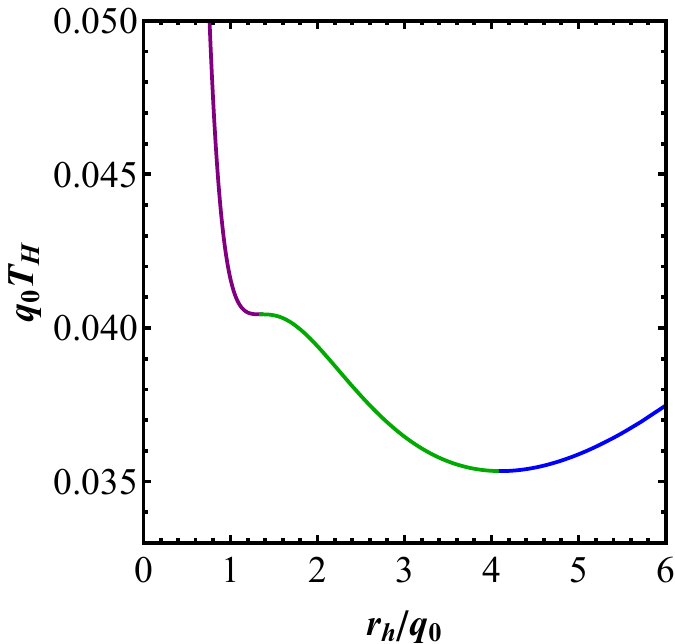} }
			\\
			\subfigure[$L \approx 7.67 q_0$, $m_0 \approx 1.39 q_0$]{\includegraphics[width=4.2cm]{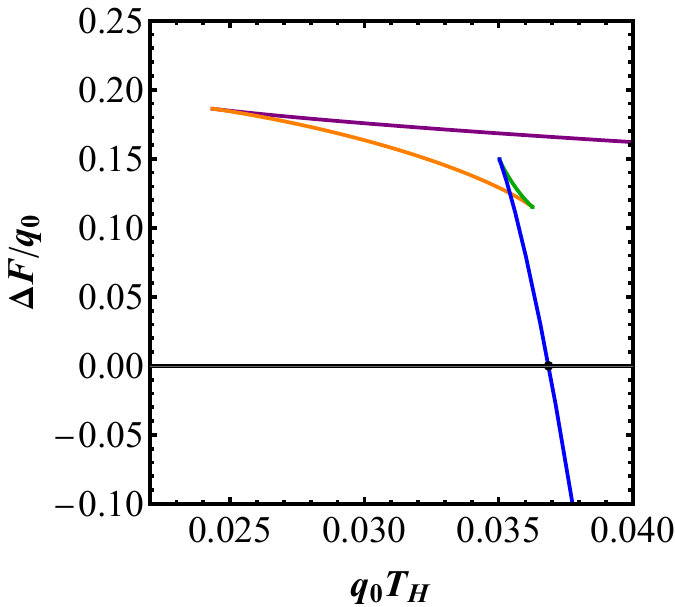}	}
			\
			\subfigure[$L \approx 7.67 q_0$, $m_0 \approx 1.85 q_0$]{\includegraphics[width=4.2cm]{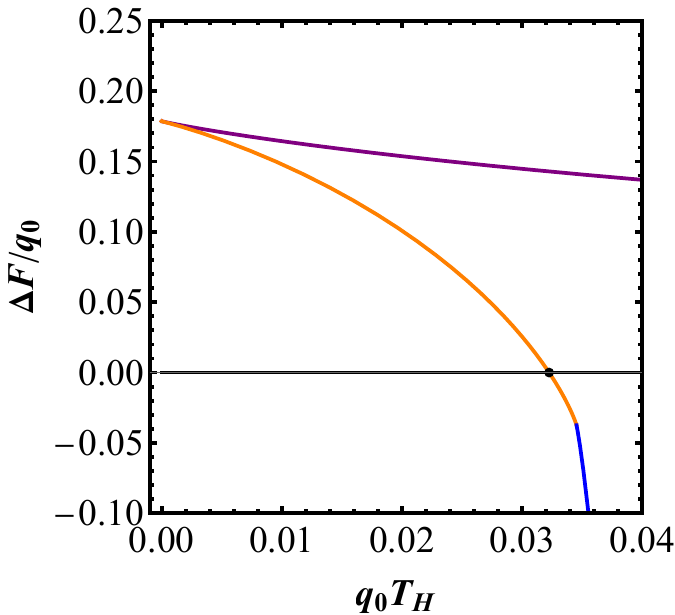} }
			\
			\subfigure[$L \approx 7.67 q_0$, $m_0 \approx 1.00 q_0$]{\includegraphics[width=4.2cm]{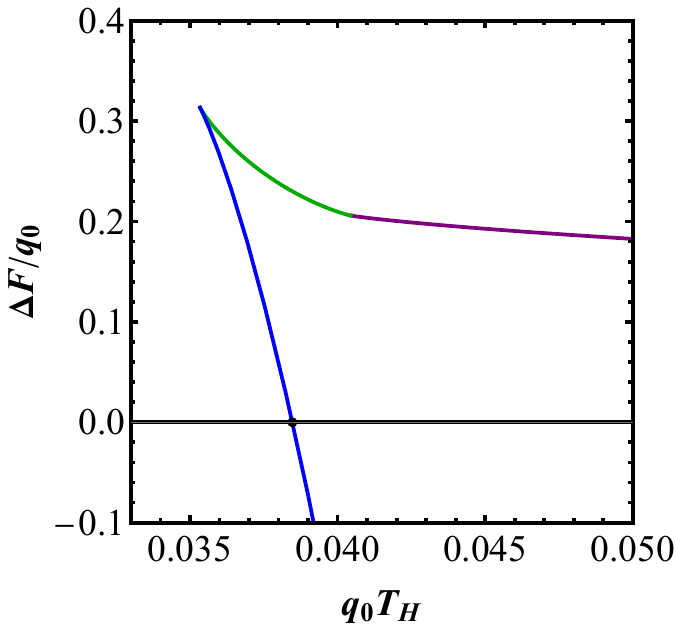} }
		\end{center}
		\caption{The characteristic curve ($r_h - T_H$ and $T_H - \Delta F$) for the Bardeen-AdS-class black hole phase states. The solid curves represent the relationship between $r_+$ and the Hawking temperature, while the dashed curves represent the $T (r_h)$ functions. The red, orange, green, and blue curves represent the tiny, small, medium, and large black hole states, respectively. The black curves represent the Bardeen-AdS state. Here, the parameters are $L \approx 7.67 q_0$ and $m_0 \approx 1.85 q_0$ satisfying Eq.~\eqref{eq_CriticalEq} and $m_0 = \hat{m}_{02}$.}
		\label{FigrTTF0m2}
	\end{figure}
\end{itemize}

From the discussions of above two cases, it emerges that in the absence of the Bardeen-AdS black hole, the Bardeen-AdS-class black hole can manifest an additional tiny black hole phase and the pure Bardeen-AdS state. This tiny black hole phase exhibits similar thermodynamical characteristics to the Schwarzschild black hole, specifically a negative heat capacity. This behavior originates from the properties of the metric \eqref{eq_GeBarBH} near $r=0$ for $m > m_0$.

\section{Conclusion and Discussion}\label{Sec_ConclusionandDiscussion}

The Bardeen black hole, a significant model of regular black holes, can be explained via nonlinear electromagnetic fields. However, its thermodynamic properties have been the subject of various controversies. To address this issue, more general black holes need to be studied, with regular black holes considered as intermediate black hole states or as pure regular spacetimes without event horizon.
	
	In our work, by using new parameters, we re-modified the action, leading to the derivation of the metric for the Bardeen-AdS-class black hole. This generalized black hole is a standard solution of nonlinear electromagnetic fields and does not involve the ``coupling constant issue", allowing for consistent construction of the thermodynamics. We focused on the ensemble with fixed $Q_m = q_0$. In this context, the Bardeen-AdS-class black hole with the parameter $m = m_0$ corresponds to a regular spacetime. Interestingly, this regular spacetime acts as an intermediate phase state of the Bardeen-AdS-class black hole in two distinct ways: one is the Bardeen-AdS black hole corresponding to the parameter range $m_0 \ge \hat{m}_{01}$, and the other is the pure Bardeen-AdS spacetime (without an event horizon) corresponding to the parameter range $0 < m_0 < \hat{m}_{01}$. As an RN gravitational system, the latter represents a novel spacetime state, bearing some resemblance to the pure AdS spacetime.
	
	Moreover, whether in the presence of the Bardeen-AdS black hole or the pure Bardeen-AdS spacetime, the thermodynamics of Bardeen-AdS-class black holes exhibits significant differences. In the former case, there are three black hole states: the large, intermediate, and small states. A phase transition occurs between the small and large black hole states, similar to that of the RN-AdS black holes. In the latter scenario, there are four black hole phase states and one regular spacetime state: the tiny, small, intermediate, and large black hole states, along with the pure Bardeen-AdS state. Besides the phase transition between the small and large black holes, there is an additional phase transition between the pure Bardeen-AdS state and black hole states. This phase transition resembles the Hawking-Page transition, which occurs between a non-black hole phase and a black hole phase. From the AdS/CFT perspective, such a phase transition may correspond to a confinement-deconfinement transition~\cite{Witten:1998Confinement}. However, when the planar and hyperbolic cases are considered, this type of phase transition does not exist. First, in the hyperbolic case, such a regular spacetime does not exist. Second, in the planar case, while a pure Bardeen-AdS state may exist under certain conditions, the free energy difference between the black hole state and the pure Bardeen-AdS state remains negative. Therefore, no phase transition occurs between the black hole and the pure Bardeen-AdS state in either the planar or hyperbolic cases. For a detailed discussion, see the Appendix~\ref{Appendix_PlanarHyperbolic}.
	
	There is another critical value, $\hat{m}_{02}$, for the Bardeen-AdS-class black holes with the pure Bardeen-AdS state. This critical value characterizes the connectedness of the domain for the radius of the outer horizon. When $\hat{m}_{02} < m_0 < \hat{m}_{01}$, the domain of $r_+$ splits into two separate regions. In this case, the $r_+ - T_H$ and $T_H - F$ curves are discontinuous, indicating that the thermodynamical topology~\cite{Wei:2022ThermodynamicDefects} of black holes may undergo transitions. When $0 < m_0 <\hat{m}_{02}$, the domain of $r_+$ remains connected. This characteristic arises from the nature of the multiple horizons in black holes and is not unique to Bardeen-AdS-class black holes.

	The Bardeen-AdS-class black hole is an extension of the Bardeen-AdS black hole, designed to ensure the naturalness and self-consistency of its thermodynamics. The presence of the Bardeen-AdS black hole as a state within the Bardeen-AdS-class black hole framework significantly affects the properties of the horizon and the thermodynamical characteristics. This underscores the importance of the Bardeen-AdS black hole, which may require further explanation and investigation.

\section*{Acknowledgments}
	
We are grateful to Prof. Yu-Xiao Liu and Dr. Hong-Yue Jiang for their valuable discussions on the Penrose diagram of the Bardeen-AdS-class black hole. This work was supported by the National Natural Science Foundation of China (Grants No. 12475055, No. 12075103, and No. 12247101).

\appendix

\section{Penrose Diagram}\label{Appendix_PenroseDiagram}
	
\begin{figure}[h]
			\begin{center}
				\subfigure[]{\includegraphics[width=5.5cm]{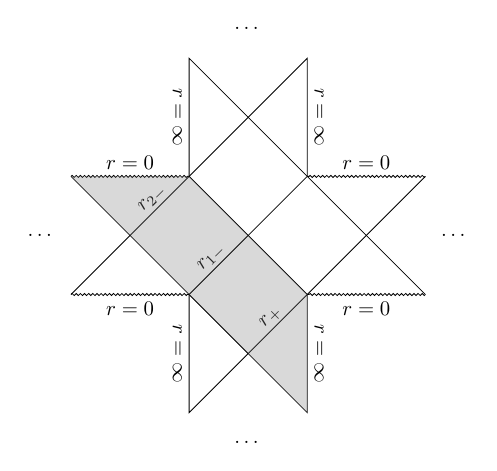}\label{FigPenroseCase1}}				
				\
				\subfigure[]{\includegraphics[width=3cm]{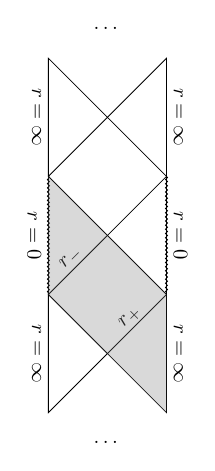} \label{FigPenroseCase2}}
				\
				\subfigure[]{\includegraphics[width=3.5cm]{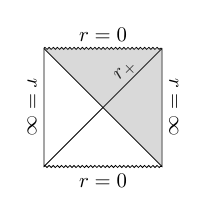} }
			\end{center}
			\caption{Penrose diagram of the non-extremal Bardeen-AdS-class black hole. Here, diagrams of non-extremal black holes under three different conditions are presented. (a), (b), and (c) represent the cases of one, two, and three horizons, respectively. Under certain conditions, the curvature singularity at $r=0$ in spacetime can be avoided in the two-horizon case. Additionally, the shaded regions represent the unextended spacetime.}\label{FigPenrose}
		\end{figure}

To further clarify and describe the spacetime structure of Bardeen-AdS-class black holes, we will present the Penrose diagram. According to the analysis in Sec.~\ref{Sec_BardeenAdSClassBHandThermodynamics}, Bardeen-AdS-class black holes can have three possible horizon configurations: one horizon, two horizons, and three horizons. First, we consider the non-extremal case. In this scenario, the function $f(r)$ in metric \eqref{eq_GeBarBH} for the three horizon configurations can be written as: for three horizons case ($(q_m / q_0) ^{3/2} < m/m_0$),
	\begin{equation}
		f\left( r \right) =N_3\left( r \right) \frac{1}{r^3}\left( r-r_+ \right) \left( r-r_{1-} \right) \left( r-r_{2-} \right);
		\label{eq_fN3}
	\end{equation}
	for two horizons case ($(q_m / q_0) ^{3/2} \ge m/m_0$),
	\begin{equation}
		f\left( r \right) =N_2\left( r \right) \frac{1}{r^2}\left( r-r_+ \right) \left( r-r_- \right);
		\label{eq_fN2}
	\end{equation}
	for one horizon case ($(q_m / q_0) ^{3/2} < m/m_0$),
	\begin{equation}
		f\left( r \right) =N_1\left( r \right) \frac{1}{r}\left( r-r_+ \right).
	\end{equation}
	The corresponding Penrose diagrams are illustrated in Fig.~\ref{FigPenrose}, respectively.
	In the cases of three horizons and one horizon, as $r$ approaches zero, $f(r)$ tends to negative infinity, indicating that the singularity at $r=0$ is space-like. In contrast, for the two-horizon case, as $r$ approaches zero, $f(r)$ tends to 1 or positive infinity, suggesting that $r=0$ is either nonsingular or a time-like singularity.

\begin{figure}[h]
			\begin{center}
				\subfigure[]{\includegraphics[width=5cm]{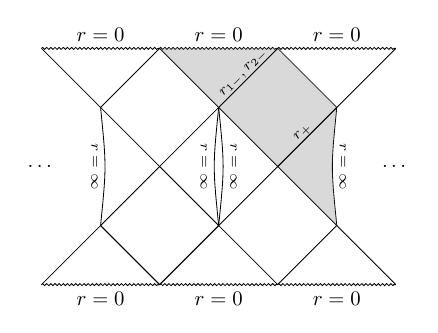}}
				\
				\subfigure[]{\includegraphics[width=4cm]{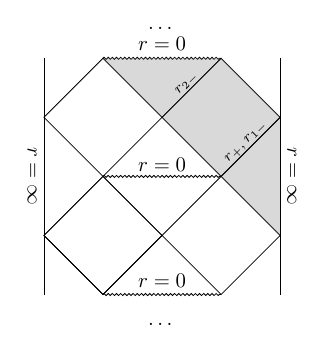} }
				\
				\subfigure[]{\includegraphics[width=3.5cm]{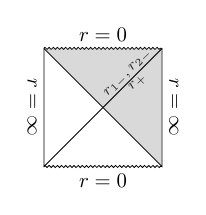} }
				\
				\subfigure[]{\includegraphics[width=2cm]{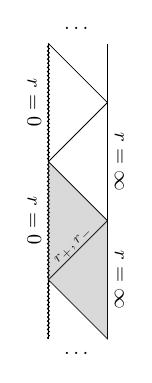} }
			\end{center}
			\caption{Penrose diagram of the extremal Bardeen-AdS-class black hole. (a), (b), and (c) represent the three extremal cases in Fig.~\ref{FigPenroseCase1} where $r_{1-} = r_{2 -}$, $r_{+} = r_{1-}$ and $ r_{+} = r_{1-} = r_{2 -} $, respectively. (d) represents the extremal case $r_+ = r_-$ in Fig.~\ref{FigPenroseCase2}. Two adjacent infinities $r =+\infty$ and $r =+\infty$ are not joined together in (a). Under certain conditions, the curvature singularity at $r=0$ in spacetime can be avoided in (d). Additionally, the shaded regions represent the unextended spacetime.}\label{FigPenroseEX}
		\end{figure}

	Secondly, for the extremal case, there are also four situations that need to be considered. Three of them correspond to the three cases for Eq.~\eqref{eq_fN3} with $r_{1-} = r_{2-}$, $r_{+} = r_{1-}$ and $r_{+} = r_{1-}= r_{2-}$, specifically,
	\begin{align}
		f\left( r \right) =N_3\left( r \right) \frac{1}{r^3}\left( r-r_+ \right) \left( r-r_{1-} \right) ^2,
		\\
		f\left( r \right) =N_3\left( r \right) \frac{1}{r^3}\left( r-r_+ \right) ^2\left( r-r_{2-} \right),
		\\
		f\left( r \right) =N_3\left( r \right) \frac{1}{r^3}\left( r-r_+ \right) ^3,
	\end{align}
	while one corresponds to a case for Eq.~\eqref{eq_fN2} with $ r_+ = r_-$, i.e.,
	\begin{equation}
		f\left( r \right) =N_2\left( r \right) \frac{1}{r^2}\left( r-r_+ \right) ^2.
	\end{equation}
	The corresponding Penrose diagrams are illustrated in Fig.~\ref{FigPenroseEX}, respectively.
	
\section{The critical point for $r_h - T(r_h)$ curve}\label{Appendix_CriticalPointForTrh}

In this appendix, we present a proposal using Fig.~\ref{FigrhT}. In the context of the Bardeen-AdS-class black hole ($q_m = q_0$ case), given that $T(r_h) = 0$ has two roots, $r_1$ and $r_2$($r_1<r_2$), the radius of horizon $r_h$, which satisfies the equations $\partial_{r_h}T(r_h) = 0$ and $\partial_{r_h}^2 T(r_h) = 0$ (see Eq.~\eqref{eq_CriticalEq}), is greater than $r_2$.

Here is the proof of this proposal. In Sec.~\ref{Sec_WithBardeenAdSBHStates}, Eq. \eqref{eq_CriticalEq} can be simplified to Eq.~\eqref{eq_CriticalEqReduce}, where we observe that the range for $r_{hc}$, the value of $r_h$ satisfying Eq.~\eqref{eq_CriticalEq}, is given by
\begin{equation}
	r_{hc}  > \sqrt{\frac{1}{24}\left(21+ \sqrt{345} \right)} q_0,
	\label{eq_r_hcL}
\end{equation}
since $m_0$ and $L$ are positive real numbers. At the critical point, the temperature is given by
\begin{equation}
	T_c =T(r_{hc}) = \frac{3r_{hc}^{2}-4 q_0^2}{2\pi r_{hc}\left( 4r_{hc}^{2}-3 q_0^2 \right)},
	\label{eq_Tc}
\end{equation}
Due to Eq.~\eqref{eq_r_hcL}, we have $T_c>0$. Moreover, through analysis, it can be determined that Eq.~\eqref{eq_rstar} provides the range of $r_1$, specifically $r_1 \le r_{*}$. By analyzing of Eq.~\eqref{eq_rstar}, we find that $r_*$ has an upper bound $\sqrt{6 } q_0/{3} $. Consequently, we have
\begin{equation}
	r_1 \le r_{*} < \frac{\sqrt{6}}{3} q_0 <\sqrt{\frac{1}{24}\left(21+ \sqrt{345} \right)} q_0 < r_{hc},
\end{equation}
which implies the root of Eq.~\eqref{eq_CriticalEq} is not in $(0, r_1)$. Moreover, considering the fact $T_c >0$, the root of Eq.~\eqref{eq_CriticalEq} is also not within $[r_1, r_2]$. Therefore, the root of Eq.~\eqref{eq_CriticalEq} can only be within $(r_2, \infty)$. This completes our proof.

Using this proposal, we can determine the monotonicity of the $r_h - T(r_h)$ curve in the interval $(0, r_1)$ shown in Fig.~\ref{FigrhT}. Considering the behavior of $T(r_h)$ at $r_h=0$ and $r=r_1$, as given by,
\begin{gather}
	\left. T\left( r_h \right) \right|_{r_h\rightarrow 0^+}=+\infty , \left. \partial _{r_h}T\left( r_h \right) \right|_{r_h\rightarrow 0^+}=-\infty ,
	\\
	\left. T\left( r_h \right) \right|_{r_h\rightarrow r_1}=0, \left. \partial _{r_h}T\left( r_h \right) \right|_{r_h\rightarrow r_1} < 0.
\end{gather}
and in conjunction with this proposal, we see that the number of stationary points of $T(r_h)$ in the interval $(0, r_1)$ remains unchanged under continuous variation of the parameters $m_0$, $q_0$, and $L$. In Fig.~\ref{Fig_Lmbyrh}, the parameter region above the green line indicates that $T(r_h)$ has two zeros. Since this parameter region is path-connected (continuous transitions between different parameter values), we can confirm the monotonicity of $T(r_h)$ in the interval $(0, r_1)$ by examining a specific point within it. Moreover, considering the monotonic behavior of $T(r_h)$ with $m=3$ and $L=5$, we can conclude that $T(r_h)$ monotonically decreases in the interval $(0, r_1)$.

\section{Other Black Hole Solutions: Planar and Hyperbolic Cases}\label{Appendix_PlanarHyperbolic}
 
In the main text, we discussed the case of spherical black holes. For general black holes, there can also be planar and hyperbolic cases. In this appendix, the solution will be written in a more general form. First, the magnetic field excited by a Dirac magnetic charge is expressed as
\begin{equation}
	A=-q_m\cos \theta d\phi ,\quad -q_mxdy,\quad -q_m\cosh \theta d\phi,
\end{equation}
for spherical, planar, hyperbolic case, respectively. Then the metric of spacetime can be obtained as
\begin{equation}
	ds^2=-f\left( r \right) dt^2+\frac{1}{f\left( r \right)}dr^2+r^2d\Omega _{k}^{2}
\end{equation}
where
\begin{gather}
	f(r)=\frac{r^2}{L^2}+k-\frac{m}{r}+\frac{1}{2r}\int_r^{\infty}{dr\left( r^2\mathcal{L} _m\left( 2q_{m}^{2}/r^4,a \right) \right)},
	\\
	d\Omega _{1}^{2}=d\theta ^2+\sin ^2\theta d\phi ^2,\quad d\Omega _{0}^{2}=dx^2+dy^2,\quad d\Omega _{-1}^{2}=d\theta ^2+\sinh ^2\theta d\phi ^2.
\end{gather}
As in most conventions, $k=1$, $0$ and $-1$ represent the spherical, planar, and hyperbolic cases, respectively. When the Lagrangian \eqref{eq_NonlinearL} is considered, the function $f(r)$ for the metric is given by
\begin{equation}
		f\left( r \right) =\frac{r^2}{L^2}+k-\frac{m}{r}+m_0\left( \frac{q_m}{q_0} \right) ^{3/2}\left( \frac{1}{r}-\frac{r^2}{\left( r^2+q_mq_0 \right) ^{3/2}} \right).
\end{equation}
Here, we briefly discuss some properties of the black hole with the fixed magnetic charge $q_m = q_0$. First, let us consider the regular spacetime with $m =m_0$, i.e.,
\begin{equation}
	f\left( r \right) =\frac{r^2}{L^2}+k-\frac{m_0 r^2}{\left( r^2+q_0^2 \right) ^{3/2}}.
\end{equation}
The existence of roots for $f(r)=0$ determines whether this regular spacetime corresponds to a black hole. Note that the asymptotic behavior
\begin{equation}
	\mathrm{for} \ r\rightarrow \infty ,\ f\left( r \right) \rightarrow \frac{r^2}{L^2};\quad \mathrm{for} \ r\rightarrow 0^+,\ f\left( r \right) \rightarrow k+\left( \frac{1}{L^2}-\frac{m_0}{q_{0}^{3}} \right) r^2.
\end{equation}
When $k = -1$, this regular spacetime must be black hole. When $k = 0$, the equation $f(r)=0$ can be further reduced as $ m_0 L^2 = {\left( r^2+q_0^2 \right) ^{3/2}}$, indicating that the existence of real positive roots requires $m_0 > q_0^3/L^2$. This indicates that, in the planar case, this regular spacetime can either represent a black hole or correspond to a pure Bardeen-AdS state without horizons. Therefore, from the perspective of regular spacetime, the properties of spherical and planar black holes are somewhat similar, both exhibiting a critical mass that characterizes the presence or absence of horizons in regular spacetime. In contrast, the regular spacetime for hyperbolic cases always possess horizons.

For further study, the Hawking temperature is a crucial and important key, and which is given by
\begin{equation}
	T_H=\frac{1}{4\pi}\left( \frac{k}{r_+}+\frac{3r_+}{L^2}-\frac{3m_0q_{0}^{2}r_+}{\left( r_{+}^{2}+q_{0}^{2} \right) ^{5/2}} \right).
\end{equation}
 Its asymptotic behavior is
 \begin{equation}
 	\mathrm{for} \ r_+\rightarrow \infty , \ T_H\rightarrow \frac{3r_+}{4\pi L^2};\quad \mathrm{for} \ r_+ \rightarrow 0^+, \ T_H \rightarrow \frac{k}{4\pi r_+}+\frac{3}{4\pi}\left( \frac{1}{L^2}-\frac{m_0}{q_{0}^{3}} \right) r_+ +\frac{15m_0r_{+}^{3}}{8\pi q_{0}^{5}}.
 	\label{eq_HawkingTHk}
 \end{equation}
For the planar case ($k = 0$),  when $m_0 > q_0^3/L^2$, there exists a positive $r_+$ such that the temperature is zero. At this point, the event horizon radius of the black hole should have a nonzero lower bound. When $m_0 \le q_0^3/L^2$, it is not difficult to see that the temperature is zero only when $r_+ = 0$, which implies that the event horizon radius of the black hole ranges from $(0, \infty)$. For the hyperbolic case ($k = -1$), the result is clear: the black hole temperature has a zero point, and $r_+$ will have a nonzero lower bound. 

Another point worth noting is whether a phase transition exists between the black hole state and the pure Bardeen-AdS state in non-spherical cases. First, for hyperbolic black holes, no pure Bardeen-AdS state exists in this system, so the notion of a phase transition does not apply. For planar black holes, under condition $m_0 \le q_0^3/L^2$, a pure Bardeen-AdS state does exist. In this case, using the on-shell Euclidean action \eqref{eq_IEuclActiononshell}, we can compute the free energy difference between the black hole state and the pure Bardeen-AdS state as follows
\begin{equation}
	\Delta F=\frac{V_2r_{+}^{3}}{16\pi}\left( -\frac{1}{L^2}-\frac{2m_0r_{+}^{2}}{\left( r_{+}^{2}+q_{0}^{2} \right) ^{5/2}}+\frac{m_0q_{0}^{2}}{\left( r_{+}^{2}+q_{0}^{2} \right) ^{5/2}} \right) , \quad V_2=\int{dx dy}.
\end{equation}
Since the temperature \eqref{eq_HawkingTHk} is greater than zero,  it can be inferred that this effective free energy is necessarily negative. This indicates that, in the planar case, although a pure Bardeen-AdS state exists, the black hole is generally more stable due to its lower free energy. As a result, no phase transition occurs between the pure Bardeen-AdS state and the black hole state.

	\bibliographystyle{unsrt}

\end{document}